\documentclass[%
 reprint,
 amsmath,amssymb,
 aps,
 showpacs,
 showkeys,
]{revtex4-2}

\usepackage{bm}
\usepackage{color}
\usepackage{subfigure}
\usepackage{amsmath}
\usepackage{graphicx}
\usepackage{dcolumn}

\begin{document}


\title{Two-color solitons in Kerr third harmonic generation model.}
\author{Alexey Sukhinin}
 \altaffiliation[]{ansukhin@umbc.edu} 
\affiliation{%
Department of Computer Science and Electrical Engineering, University of Maryland, Baltimore County, MD, 21250, USA}
\author{Natalia Litchinitser}
\affiliation{Department of Electrical and Computer Engineering, Duke University, Durham, NC, 27708, USA}
 \author{Curtis Menyuk}
\affiliation{Department of Computer Science and Electrical Engineering, University of Maryland, Baltimore County, MD, 21250, USA}
\author{Jean-Claude Diels}
\affiliation{Department of Electrical and Computer Engineering, University of New Mexico, Albuquerque, NM, 87131, USA}
\author{Alejandro B Aceves}
\affiliation{%
 Department of Mathematics, Southern Methodist University, Dallas, TX, 75275, USA}

\begin{abstract}

We investigate two-color, two dimensional spatially localized light modes in a resonant Kerr third-harmonic generation model. Using computational tools, we identify two distinct families of localized states. Unlike the single-component 2D nonlinear Schrödinger equation and previously studied non-resonant two-color systems, the dynamics are not dictated by a universal critical power, but depends on the distribution of power between the harmonics. The "fundamental-dominated" family acts as a "dynamical separatrix" between simultaneous collapse and joint diffraction, whereas the "third-harmonic-dominated" family does not. We further identify resonant collapse events accompanied by strong oscillations of the third harmonic, revealing a collapse mechanism absent from standard Kerr self-focusing.
\end{abstract}

\pacs{42.65.Jx, 42.65.Sf, 42.65.Tg, 52.38.Hb, 52.35.Sb}

\keywords{Self-focusing, Kerr nonlinearity, Solitons, Cross-phase modulation, Third harmonic generation}
\maketitle

Collapse of optical beams in Kerr media is a paradigmatic nonlinear phenomenon governed by the (2+1)D nonlinear Schrödinger equation (NLSE), where self-focusing above a critical power leads to singular behavior and universal approach to the Townes soliton \cite{chiao1964self,gaeta2000catastrophic,moll2003self,chen2020observation,bakkali2021realization}. While this framework is well established for single-component fields, its extension to multi-frequency systems remains incomplete.

Motivated by applications in laser filamentation, frequency conversion, and broadband coherent light generation, significant effort has been devoted to multi-color nonlinear interactions \cite{akozbek2002third, berge20133d,fedorov2018extreme}. In particular, two-color self-focusing dynamics have been explored in non-resonant configurations, where coupling arises through cross-phase modulation, leading to simultaneous collapse and generalized power thresholds \cite{sukhinin2017collapse,sukhinin2019two}. 

In contrast, resonant interactions-such as third-harmonic generation (THG), where $\omega_2=3\omega_1$-introduce coherent energy exchange that qualitatively modifies the nonlinear dynamics. Although harmonic generation and multi-wave coupling have been extensively studied in structured and cavity-based systems, the role of resonant coupling in spatial self-focusing and collapse remains largely unexplored. More broadly, the present problem raises the question of whether resonantly coupled multicolor beams exhibit self-focusing dynamics beyond the scope of the classical Marburger formula for the collapse distance \cite{marburger1975self}, motivating the development of generalized scaling laws for multicolor self-focusing.

Here we show that resonant Kerr coupling leads to fundamentally new self-focusing behavior that cannot be reduced to either the standard NLSE or previously studied two-color models. We consider a third-harmonic generation system described by coupled (2+1)D nonlinear Schrödinger equations with both cubic Kerr nonlinearity and resonant interaction terms.

Based on a coupled 2-D NLSE model, we demonstrate the existence of two families of two-color localized states, and classified them according to which harmonic carries the larger fraction of the total power. Remarkably, we find that the concept of a universal critical power breaks down; in contrast to the single-component NLSE and non-resonant two-color systems, no global collapse threshold exists. Instead, the dynamics are governed by the distribution of power between the harmonics. For the family in which the fundamental component carries the larger share of the power, the localized states act as "dynamical separatrices" between simultaneous collapse and joint diffraction. Furthermore, we identify strongly oscillatory behavior of the third-harmonic component as a precursor to resonant collapse, revealing a collapse mechanism absent from previously studied models. In contrast, no analogous separatrix behavior is observed for the family in which the third-harmonic component carries the larger share of the power. 

These results establish a new regime of multi-frequency self-focusing in Kerr media and challenge the universality of collapse thresholds, with direct implications for exploration of filament formation and nonlinear wave dynamics in resonant systems.

We start by considering the quasi-monochromatic approximation 
\[E = E_1e^{-i\omega_1 t}+E_2e^{-i\omega_2 t} + c.c\]
where $\omega_1$ and $\omega_2$ are the carrier frequencies and $E_1, E_2$ being the corresponding envelopes.
In the resonant case, we assume
\[\omega_2=3\omega_1\,\,\,(k_2 \approx 3k_1)\]
giving the following relevant terms from the cubic nonlinearity
\begin{eqnarray} \nonumber
|E|^2E= 3|E_1|^2E_1e^{-i\omega_1 t}+6|E_2|^2E_1e^{-i\omega_1 t}+\\ \nonumber
3E_2(E_1^*)^2e^{2i \omega_1 -i\omega_2}+3|E_2|^2E_2e^{-i\omega_2 t}+ \\ \nonumber
6|E_1|^2E_2e^{-i\omega_2 t}+ E_1^3e^{-3i\omega_1}+...
\end{eqnarray}

After combining terms with the same frequencies, the model becomes

\begin{eqnarray} \nonumber
\frac{2}{3}k_0n_0i\frac{\partial E_{1}}{\partial z} +  \Delta_{\bot}E_{1} + \frac{3}{9}n_2n_0k_0^2(|E_{1}|^2 + 2|E_{2}|^2)E_{1}+ \\ \nonumber
\frac{3}{9}n_2n_0k_0^2E_2(E_1^*)^2  = 0\,\,\,\,\, \\ \nonumber
2k_0n_0i\frac{\partial E_{2}}{\partial z} + \Delta_{\bot}E_{2} + 3n_2n_0k_0^2(2|E_{1}|^2 + |E_{2}|^2)E_{2} + \\ \nonumber
n_2n_0k_0^2E_1^3 = 0\,\,\,\,\,
\end{eqnarray}
where $E_i=E_i(x,y,z)$, $\Delta_{\bot}=\frac{\partial^2 }{\partial x^2}+\frac{\partial^2}{\partial y^2}$.\\
\\
Making the equations non dimensional,
\[z=z_0z,\,\,\,\,x=r_0x,\,\,\,\,y=r_0y,\,\,\,\,E_1=E_0E_1,\,\,\,\,E_2=E_0E_2\]
where
\begin{eqnarray} \nonumber
r_0=\sqrt{\frac{k_0n_0z_0}{2}},\,\,\,\ E_0=\sqrt{\frac{1}{3n_2r_0^2}}
\end{eqnarray} 
we arrive at
\begin{eqnarray} \label{main1} 
i\frac{\partial E_{1}}{\partial z} +  3\Delta_{\bot}E_{1} + \frac{1}{3}(|E_{1}|^2 + 2|E_{2}|^2)E_{1}+\frac{1}{3}E_2(E_1^*)^2  = 0\,\,\,\\ 
\label{main2}
i\frac{\partial E_{2}}{\partial z} + \Delta_{\bot}E_{2} +(2|E_{1}|^2 + |E_{2}|^2)E_{2} + \frac{1}{3}E_1^3 = 0\,\,\,
\end{eqnarray}

\subsection*{Search for Resonant Critical Power}

Collapse of a single (quasi-monochromatic) Gaussian beam depends on the initial conditions, in particular cross-sectional power. Critical power for collapse is given by $P_{cr}= \alpha \lambda^2/(4 \pi n_0 n_2)$, where $\lambda$ the wavelength and the power of the Townes soliton is equal to $\alpha$, i.e.
\[\alpha =\int |{\cal E}(r)|^2rdr \approx 1.86225\]
where ${\cal E}(r)$ is the radially symmetric ground state solution of the nonlinear boundary value problem $-\beta{\cal E}+\Delta_{\bot}{\cal E}+{\cal E}^3=0$, with $\partial {\cal E}/{\partial r}(0)=0$, and ${\cal E}(\infty)=0$, for $\beta=1$, which can be solved numerically.
 
Invariant scaling ${\cal E}(r) = \sqrt{\beta}{\cal E}(s)$, $s= \sqrt{\beta} r$ can eliminate $\beta$ in the above mentioned equation. This means that the amplitude and the width of the Townes soliton can be varied by $\beta$ while maintaining the same power. One of the most relevant results is the fact that collapsing gaussian beam converges to the Townes profile for $\beta\rightarrow \infty$.
 
In the case of two-color non-resonant collapse, there is a critical power criterion, which effectively splits into the powers of the individual beams. In \cite{sukhinin2017collapse} it was found that two-color beams with initial power slightly above total critical, undergo simultaneous collapse if their profiles are close to the corresponding two-color soliton. Furthermore, there is no  scenario under which one beam collapses while the other diffracts.

As we will now see, the dynamics in  the resonant case is more complex. We first seek for stationary solutions  
\[E_{1}(x,y,z) = {\cal E}_{1}(x,y)e^{i\beta_{1} z },\,\,\,\, E_{2}(x,y,z) = {\cal E}_{2}(x,y)e^{i\beta_{2} z }\]
where $\beta_{2}=3\beta_{1}=\beta$ are the propagation constants. This results in the system of nonlinear equations
\begin{eqnarray}
\label{bvp1}
-{\cal E}_{1} +  9\Delta_{\bot}{\cal E}_{1} + ({\cal E}_{1}^2 + 2{\cal E}_{2}^2){\cal E}_{1}+{\cal E}_{2}{\cal E}_{1}^{2}  = 0 \\ 
\label{bvp2}
-{\cal E}_{2} + \Delta_{\bot}{\cal E}_{2} +(2{\cal E}_{1}^2 + {\cal E}_{2}^2){\cal E}_{2} + \frac{1}{3}{\cal E}_1^3 = 0
\end{eqnarray}
where the parameter $\beta$ is eliminated when we made the  substitution ${\cal E}_i(r) = \sqrt{\beta}{\cal E}_i(s)$, $s= \sqrt{\beta} r$.

\begin{figure}[htbp]
  \begin{center}
    \mbox{
      \subfigure[]{\scalebox{0.5}{\includegraphics[width=3in,height=2in]{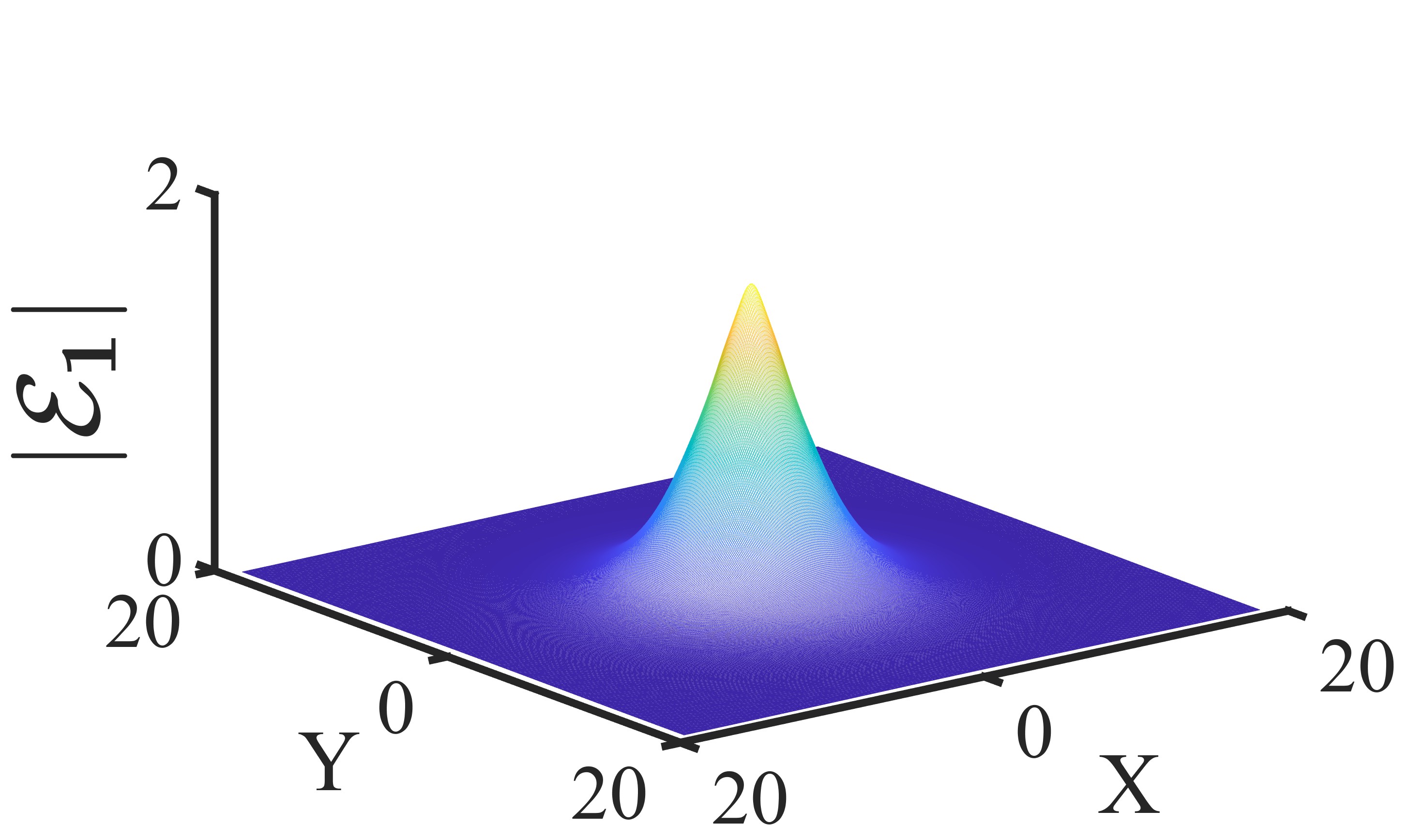}}}} \quad
      \subfigure[]{\scalebox{0.5}
{\includegraphics[width=3in,height=2in]{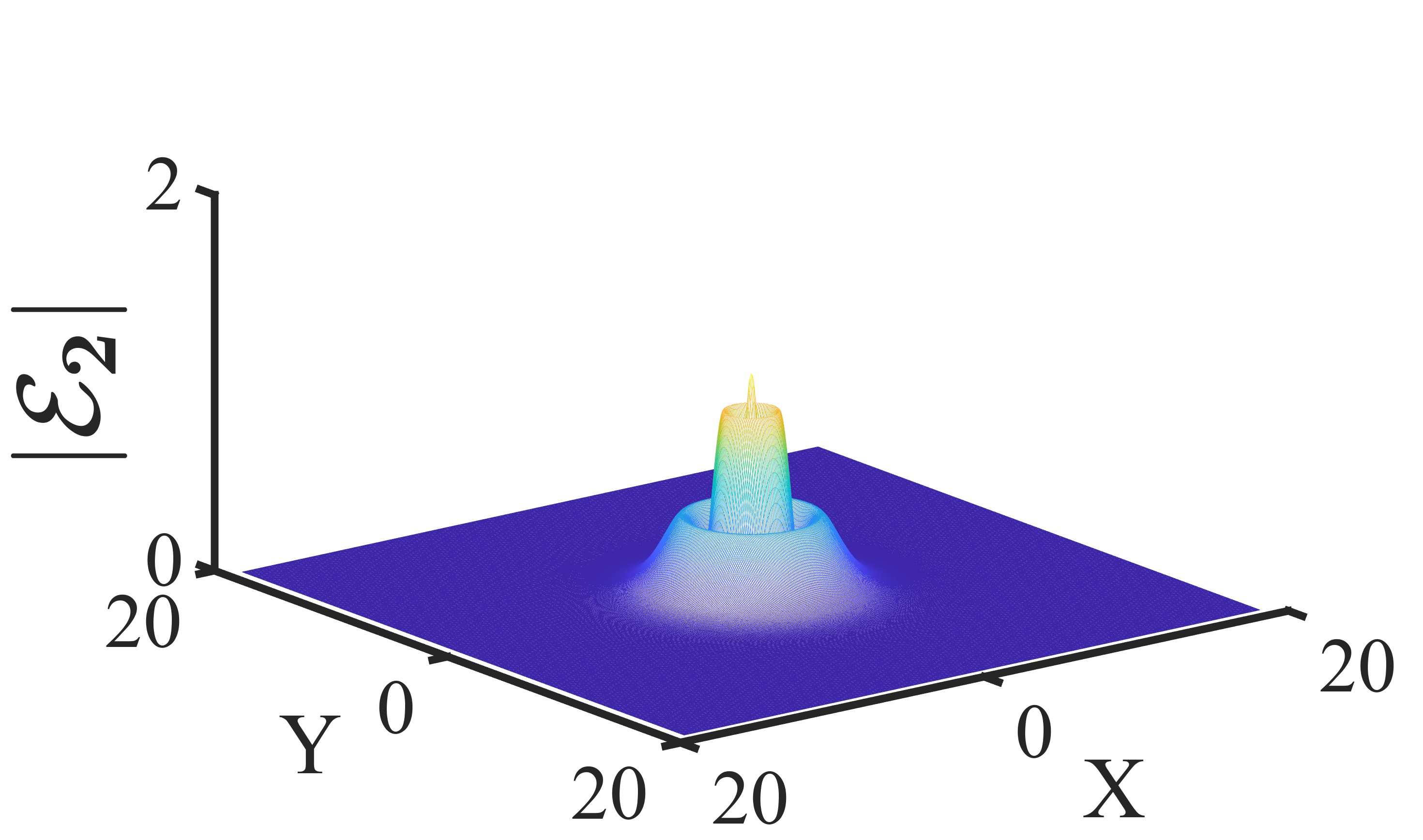}}}
      \subfigure[]{\scalebox{0.5}
{\includegraphics[width=3in,height=2in]{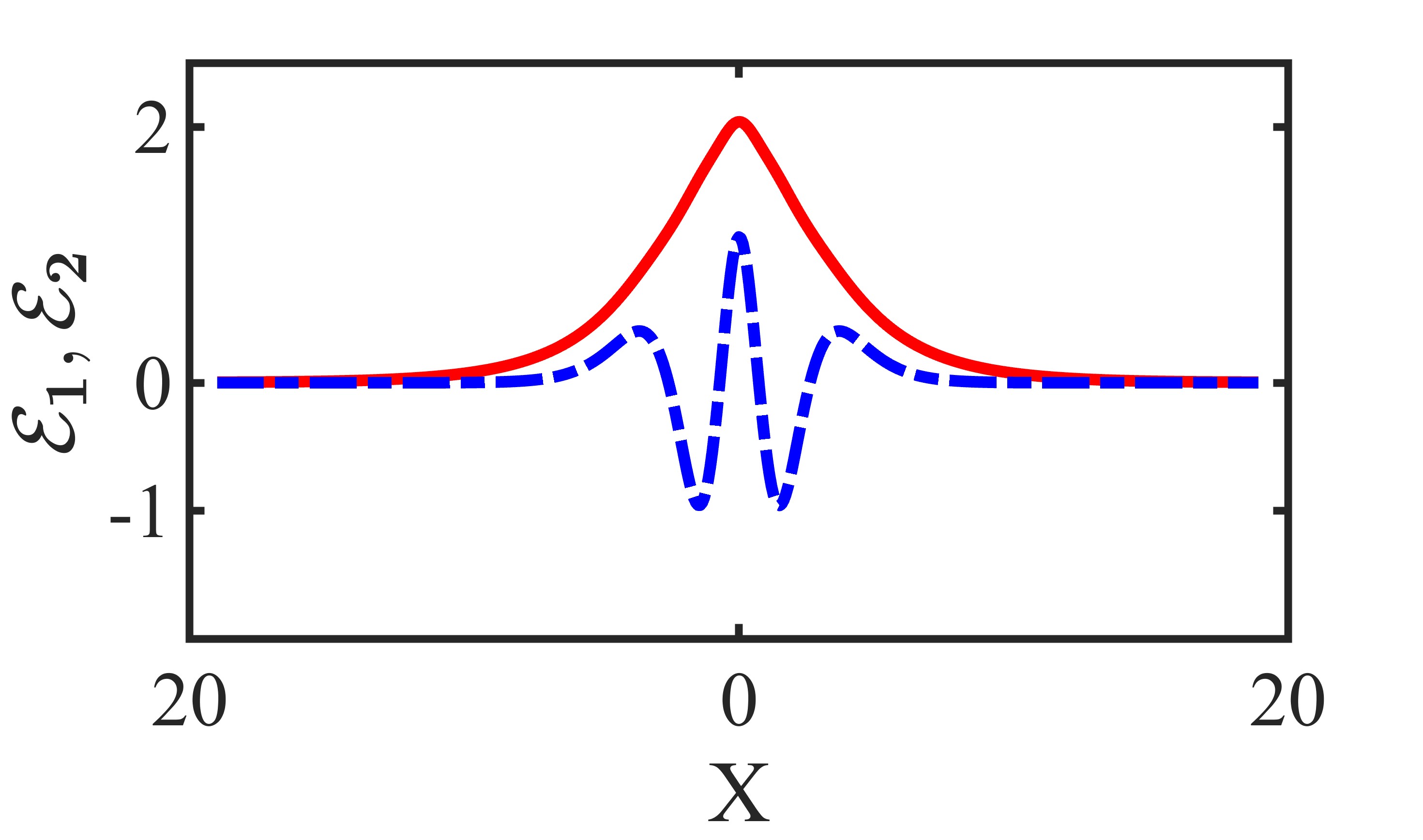}}}
	  \subfigure[]{\scalebox{0.5}
{\includegraphics[width=3in,height=2in]{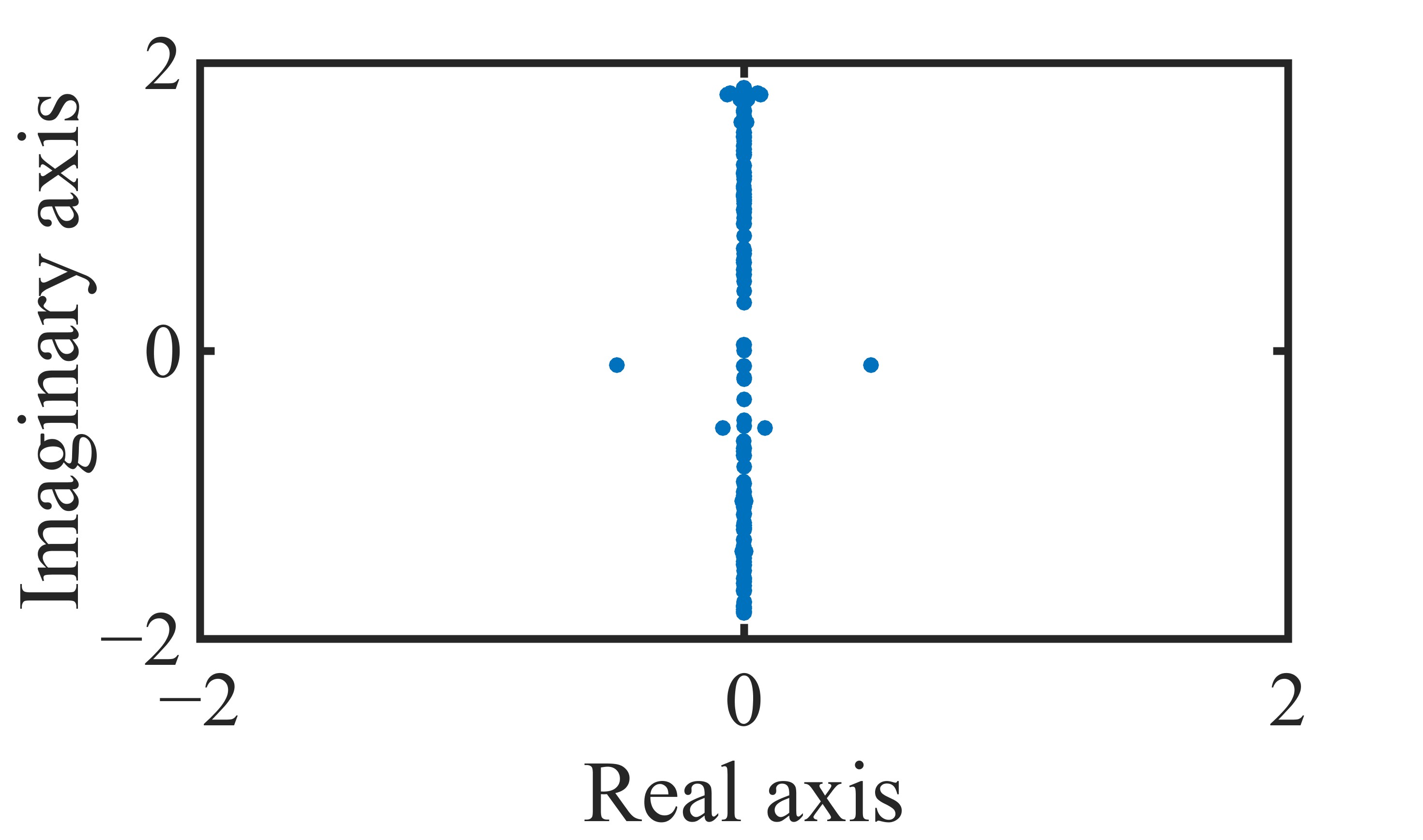}}} \quad
    \caption{(a) 3D profile of field $|{\cal E}_1(x,y)|$ with power  $P_1=15.9753$. (b) 3D profile of field $|{\cal E}_2(x,y)|$with power $P_2=2.3873$. (c) Radial profile of both electric fields ${\cal E}_1(x,y)$ (red) and ${\cal E}_2(x,y)$ (blue) with total power $P=P_1+P_2=18.3626$. (d) Linear stability spectrum}
    \label{soliton1}
  \end{center}
\end{figure}

\begin{figure}[htbp]
  \begin{center}
    \mbox{
      \subfigure[]{\scalebox{0.5}{\includegraphics[width=3in,height=2in]{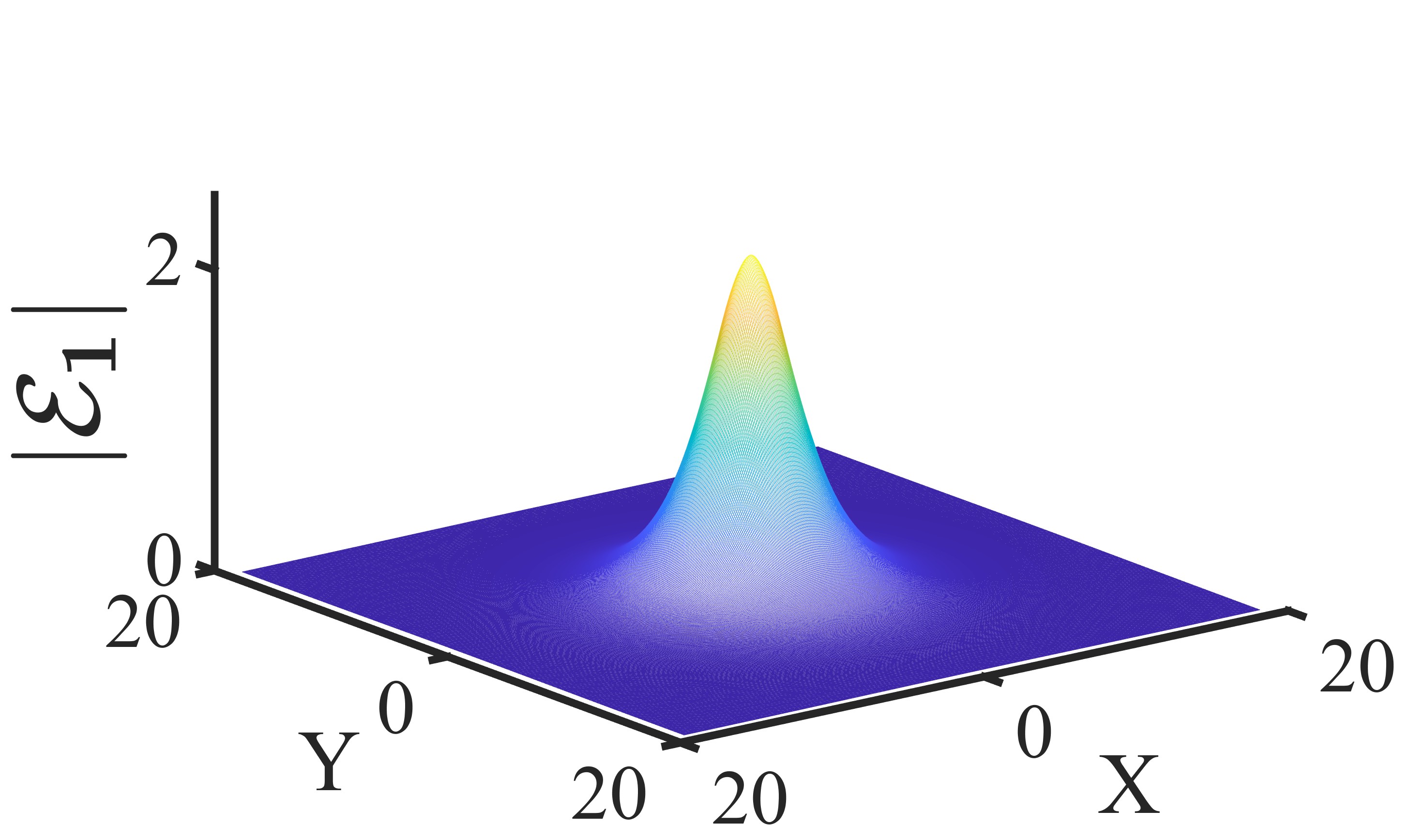}}}} \quad
      \subfigure[]{\scalebox{0.5}
{\includegraphics[width=3in,height=2in]{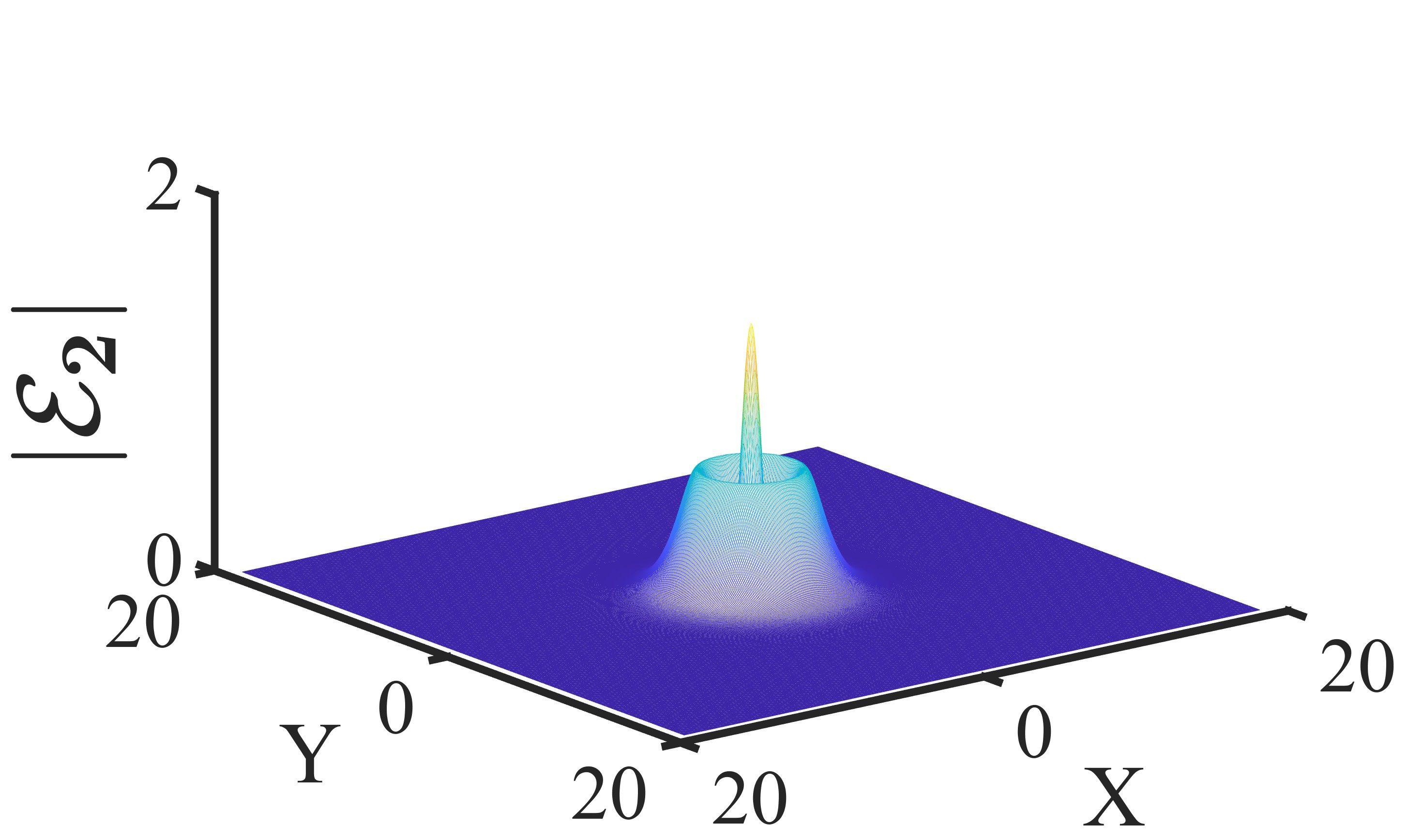}}}
      \subfigure[]{\scalebox{0.5}
{\includegraphics[width=3in,height=2in]{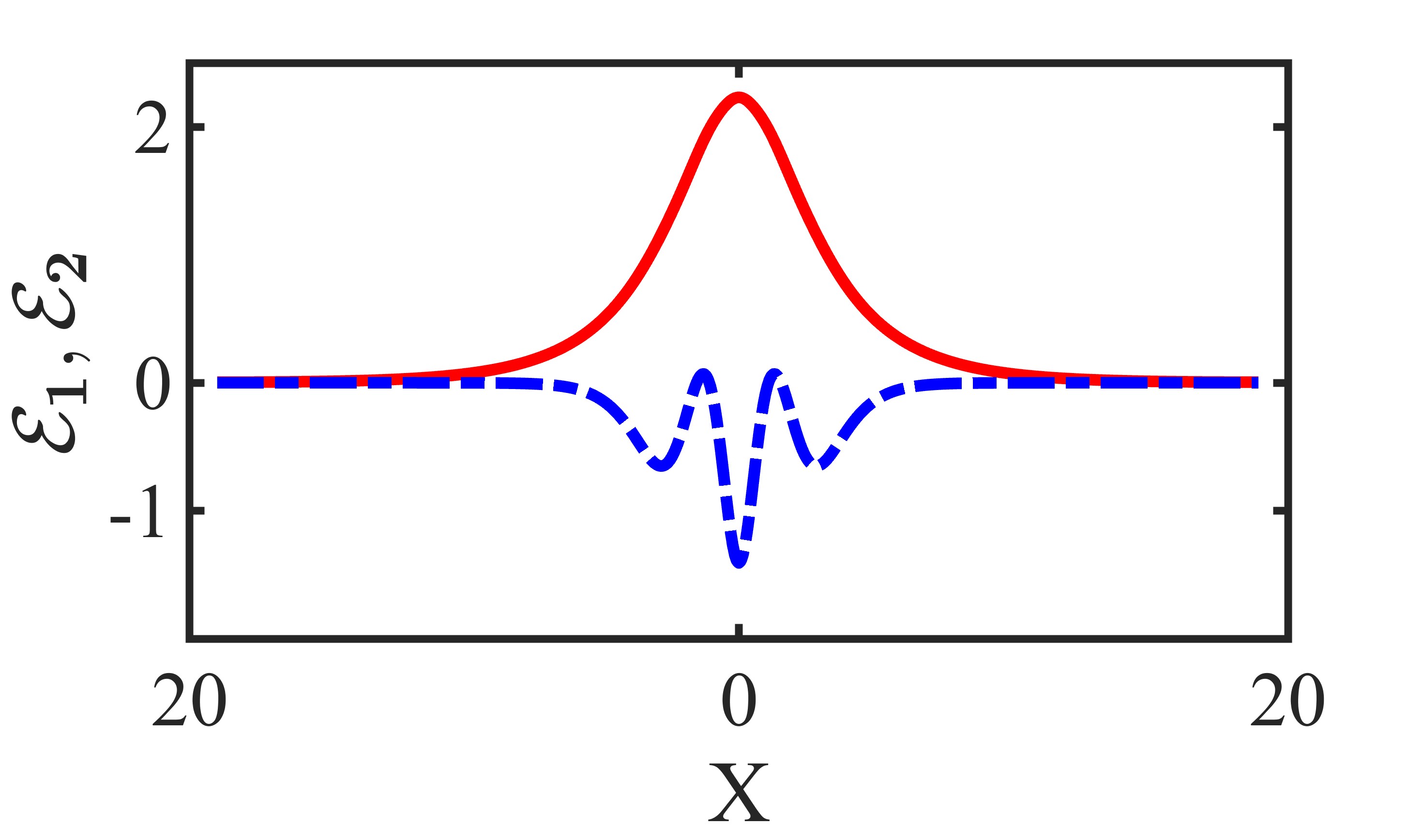}}}
	  \subfigure[]{\scalebox{0.5}
{\includegraphics[width=3in,height=2in]{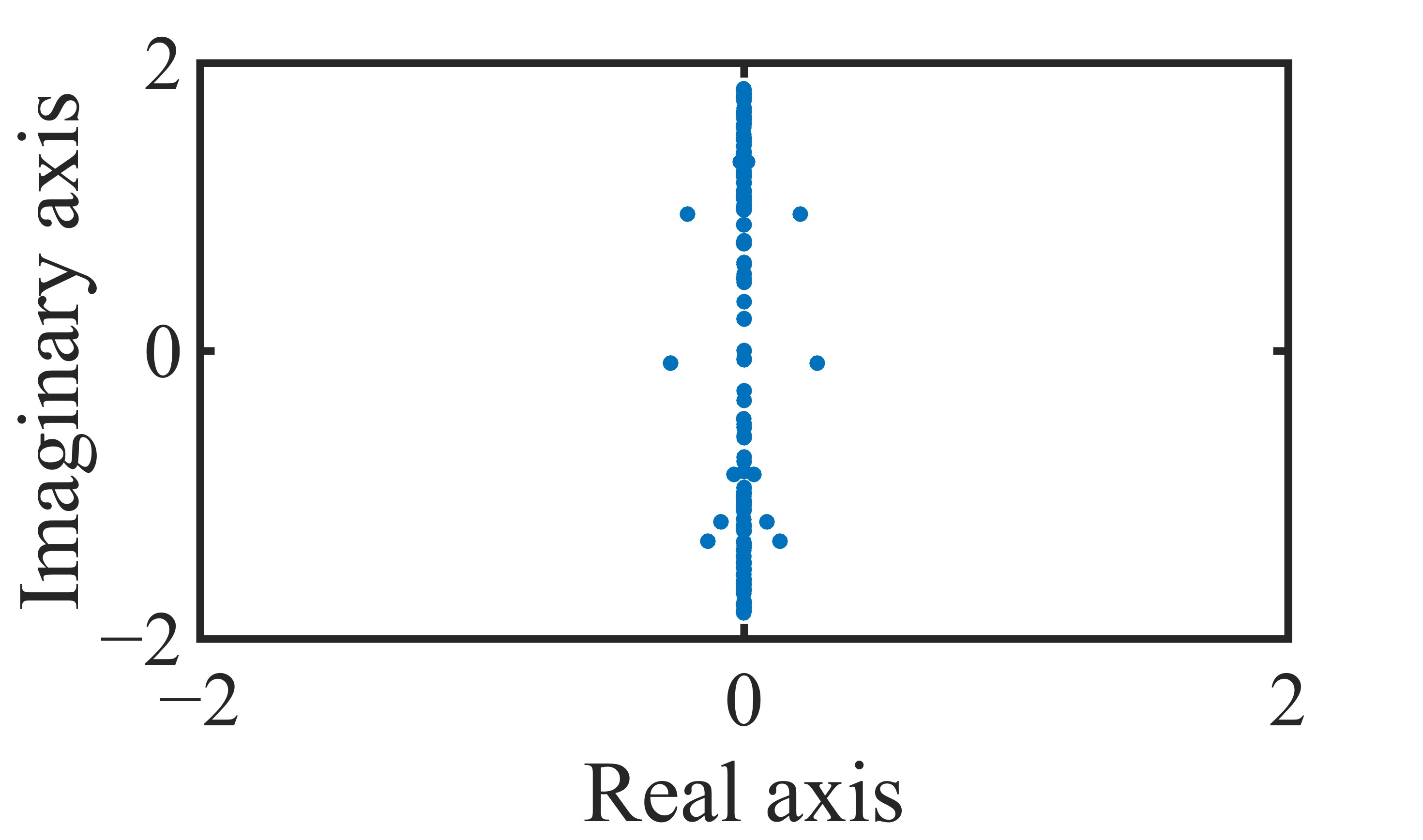}}} \quad
    \caption{3D profile of field $|{\cal E}_1(x,y)|$ with power  $P_1=17.4778$. (b) 3D profile of field $|{\cal E}_2(x,y)|$with power $P_2=2.2858$. (c) Radial profile of both electric fields ${\cal E}_1(x,y)$ (red) and ${\cal E}_2(x,y)$ (blue) with total power $P=P_1+P_2=19.7636$. (d) Linear stability spectrum}
    \label{soliton2}
  \end{center}
\end{figure}

\begin{figure}[htbp]
  \begin{center}
    \mbox{
      \subfigure[]{\scalebox{0.5}{\includegraphics[width=3in,height=2in]{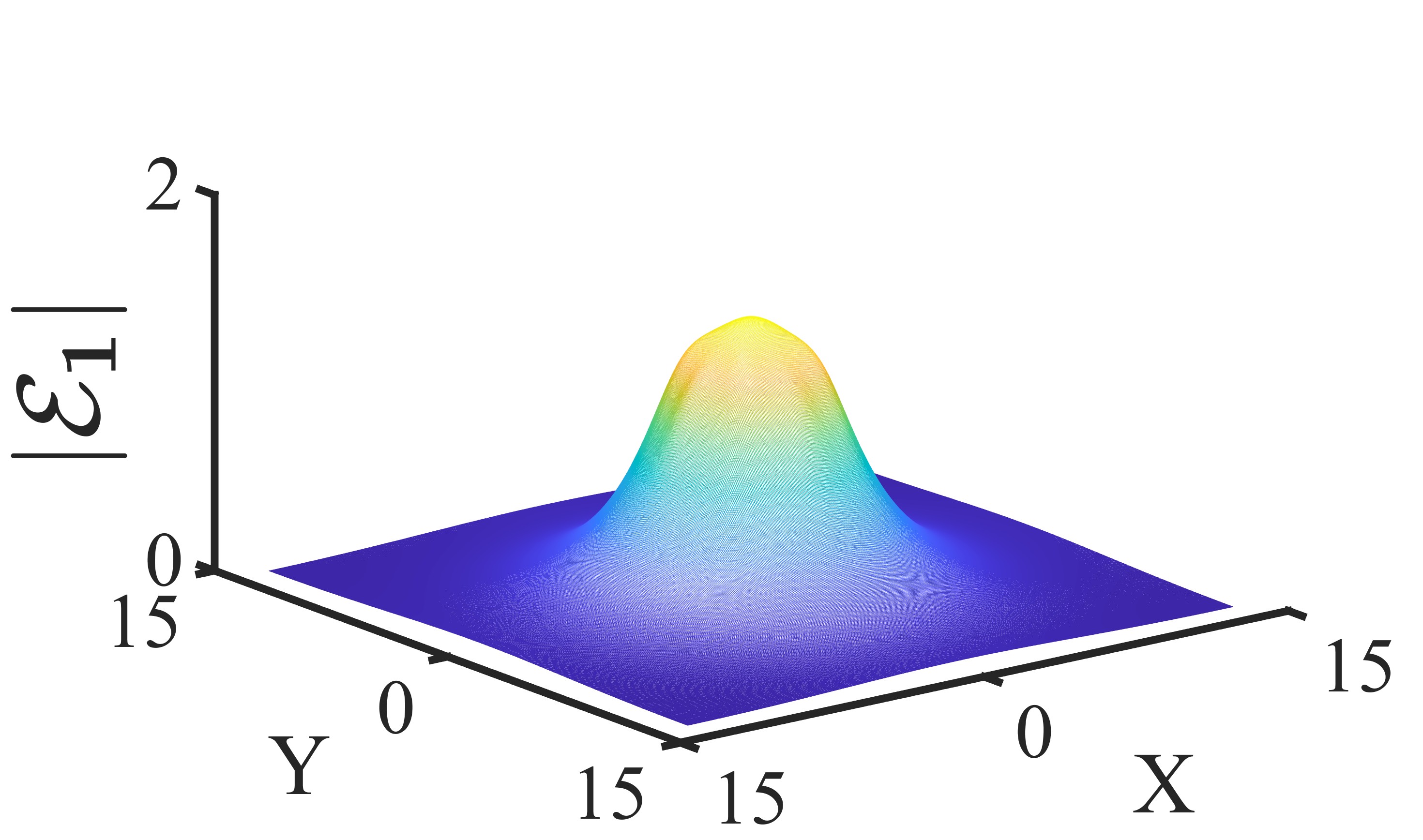}}}} \quad
      \subfigure[]{\scalebox{0.5}
{\includegraphics[width=3in,height=2in]{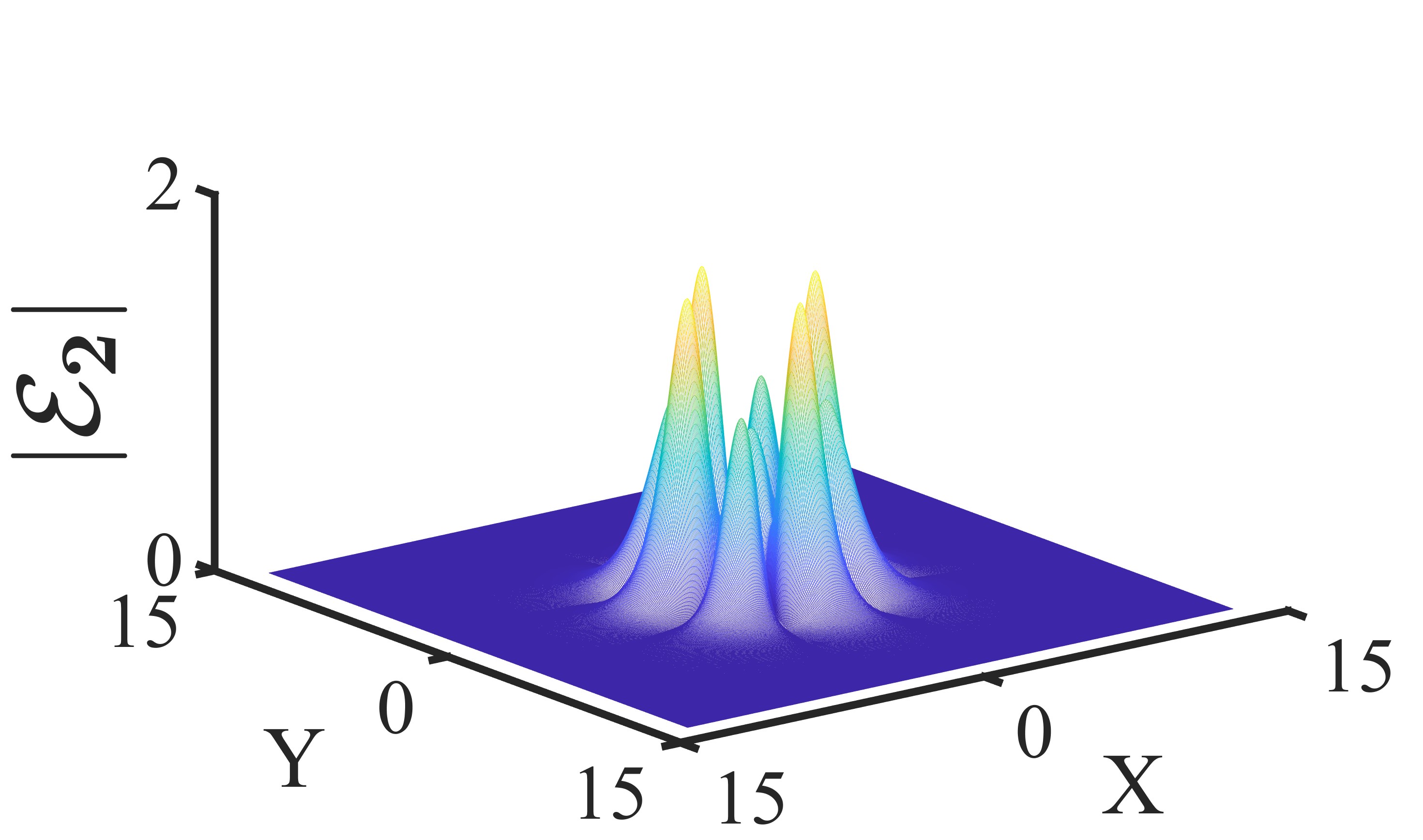}}}
      \subfigure[]{\scalebox{0.5}
{\includegraphics[width=3in,height=2in]{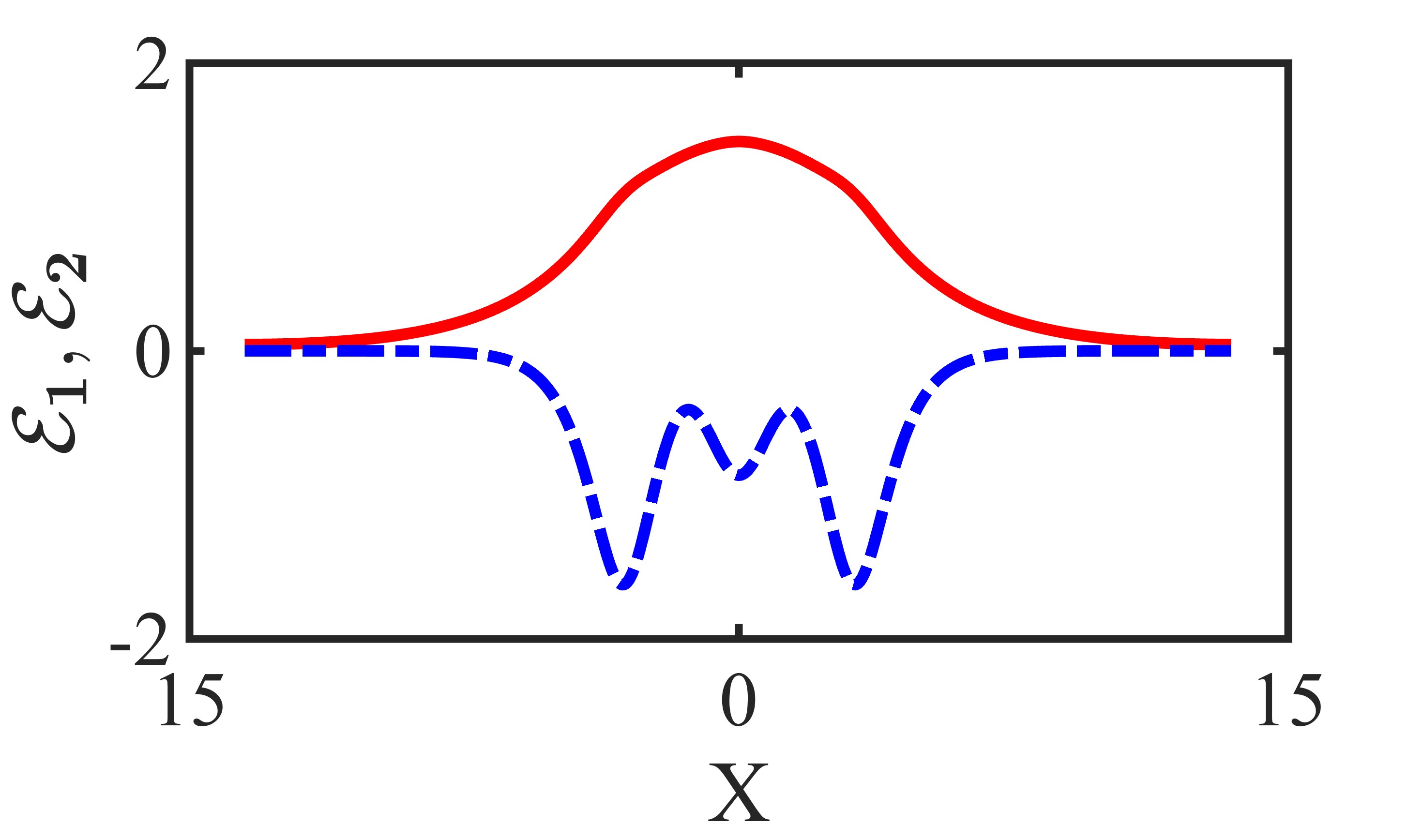}}}
	  \subfigure[]{\scalebox{0.5}
{\includegraphics[width=3in,height=2in]{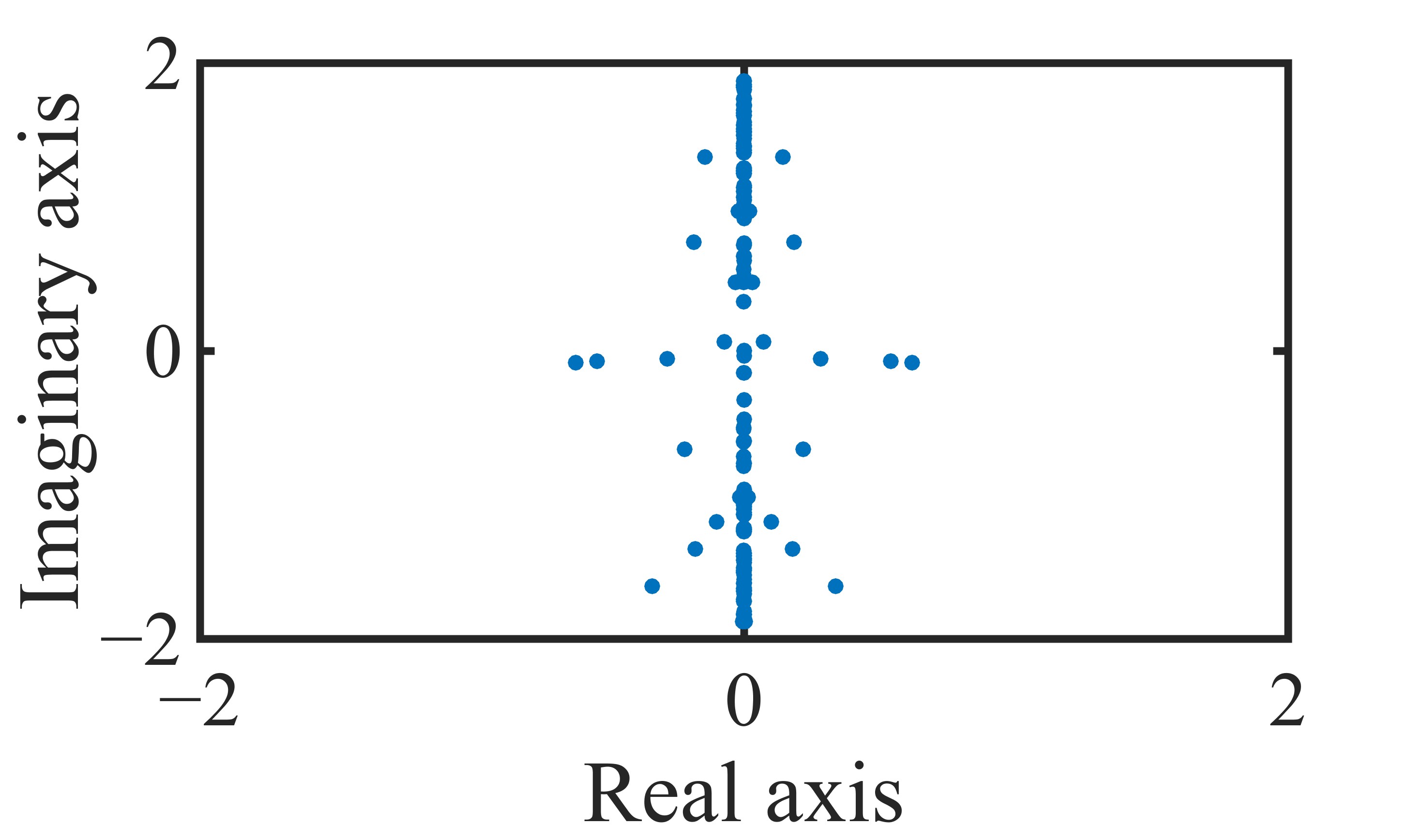}}} \quad
    \caption{3D profile of field $|{\cal E}_1(x,y)|$ with power  $P_1=15.6411$. (b) 3D profile of field $|{\cal E}_2(x,y)|$with power $P_2=4.8127$. (c) Radial profile of both electric fields ${\cal E}_1(x,y)$ (red) and ${\cal E}_2(x,y)$ (blue) with total power $P=P_1+P_2=20.4538$. (d) Linear stability spectrum}   
    \label{soliton3}
  \end{center}
\end{figure}

\begin{figure}[htbp]
  \begin{center}
    \mbox{
      \subfigure[]{\scalebox{0.5}{\includegraphics[width=3in,height=2in]{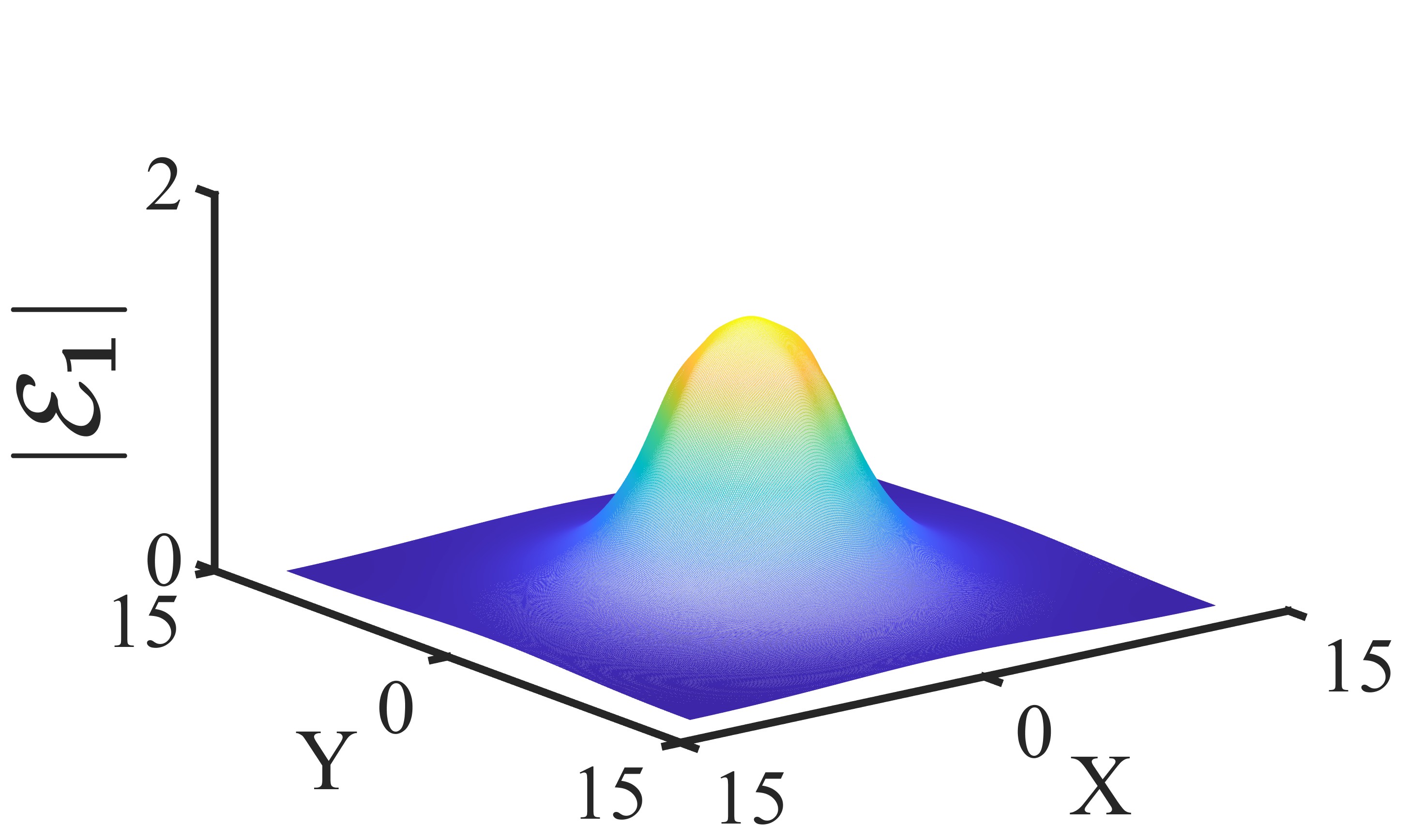}}}} \quad
      \subfigure[]{\scalebox{0.5}
{\includegraphics[width=3in,height=2in]{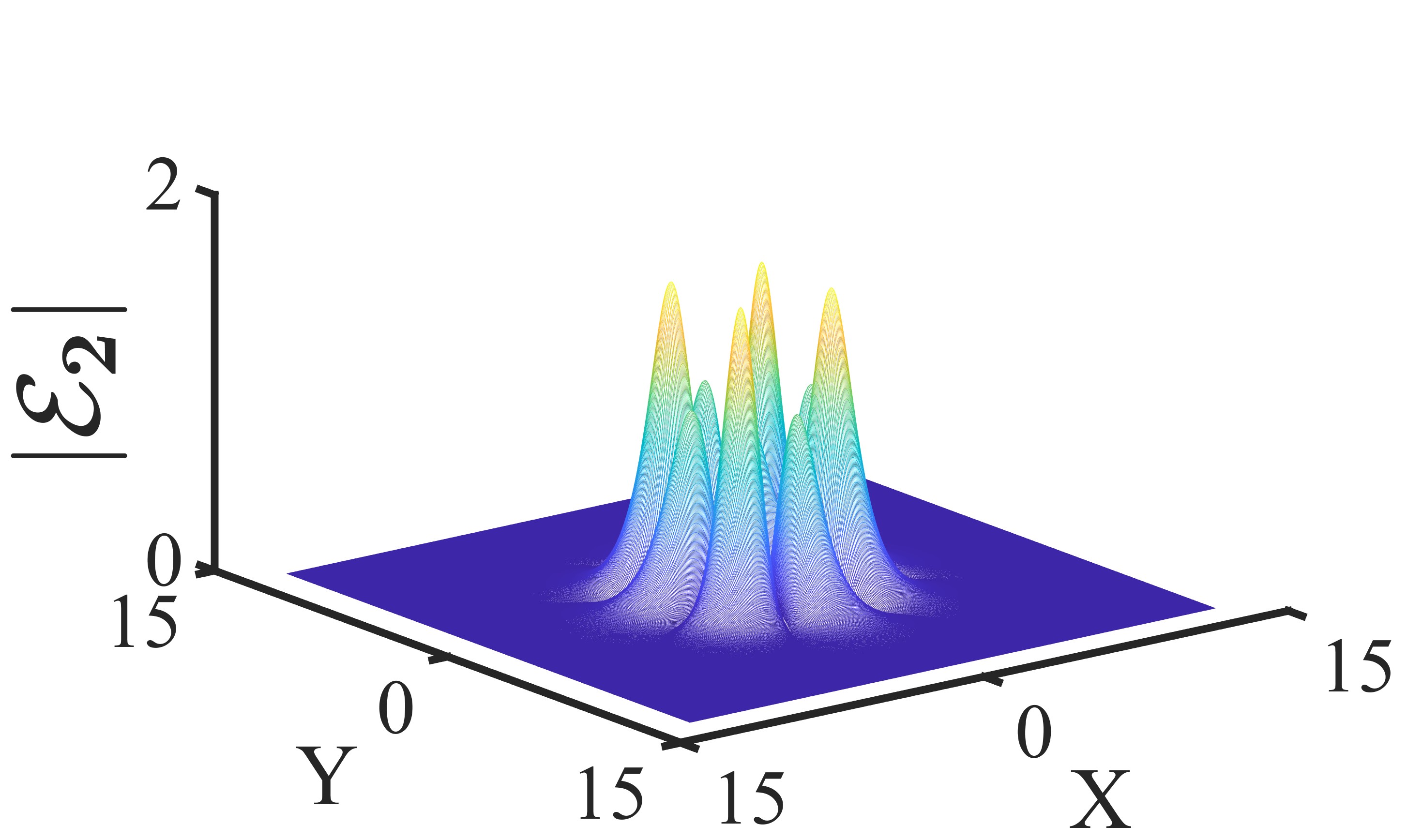}}}
      \subfigure[]{\scalebox{0.5}
{\includegraphics[width=3in,height=2in]{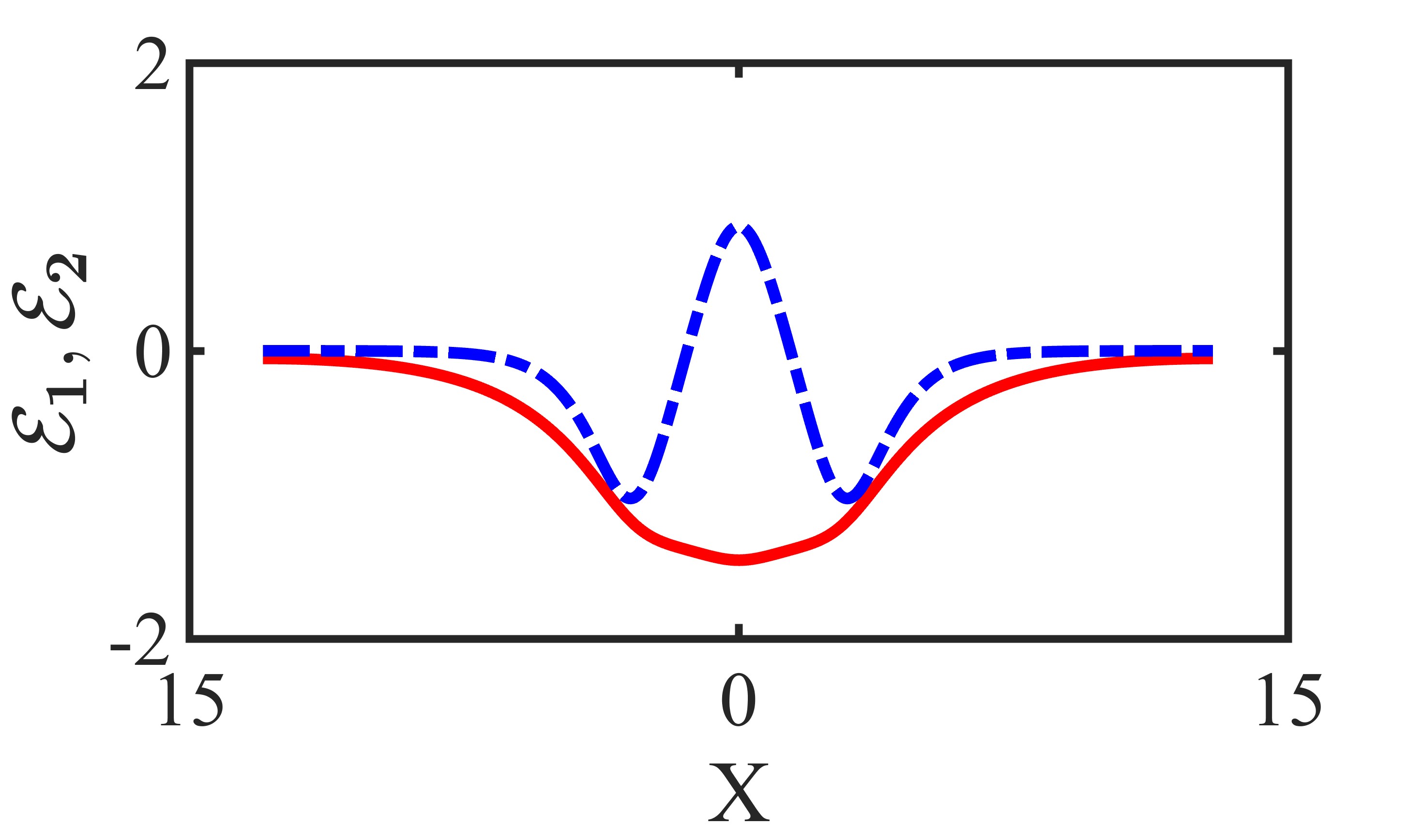}}}
	  \subfigure[]{\scalebox{0.5}
{\includegraphics[width=3in,height=2in]{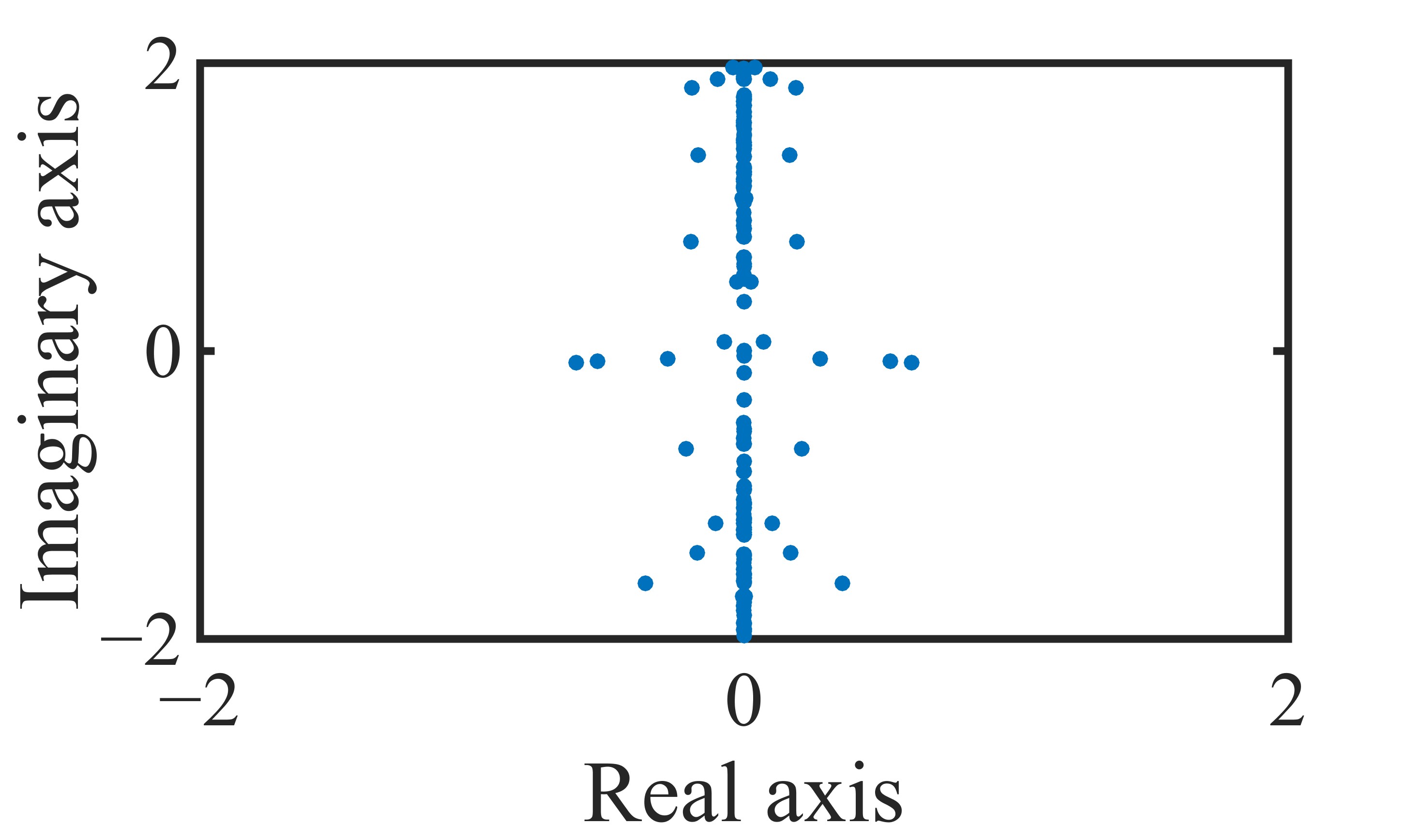}}} \quad
    \caption{3D profile of field $|{\cal E}_1(x,y)|$ with power  $P_1=15.6428$. (b) 3D profile of field $|{\cal E}_2(x,y)|$ with power $P_2=4.8120$. (c) Radial profile of both electric fields ${\cal E}_1(x,y)$ (red) and ${\cal E}_2(x,y)$ (blue) with total power $P=P_1+P_2=20.4548$. (d) Linear stability spectrum}    
    \label{soliton4}
  \end{center}
\end{figure}

\begin{figure}[htbp]
  \begin{center}
    \mbox{
      \subfigure[]{\scalebox{0.5}{\includegraphics[width=3in,height=2in]{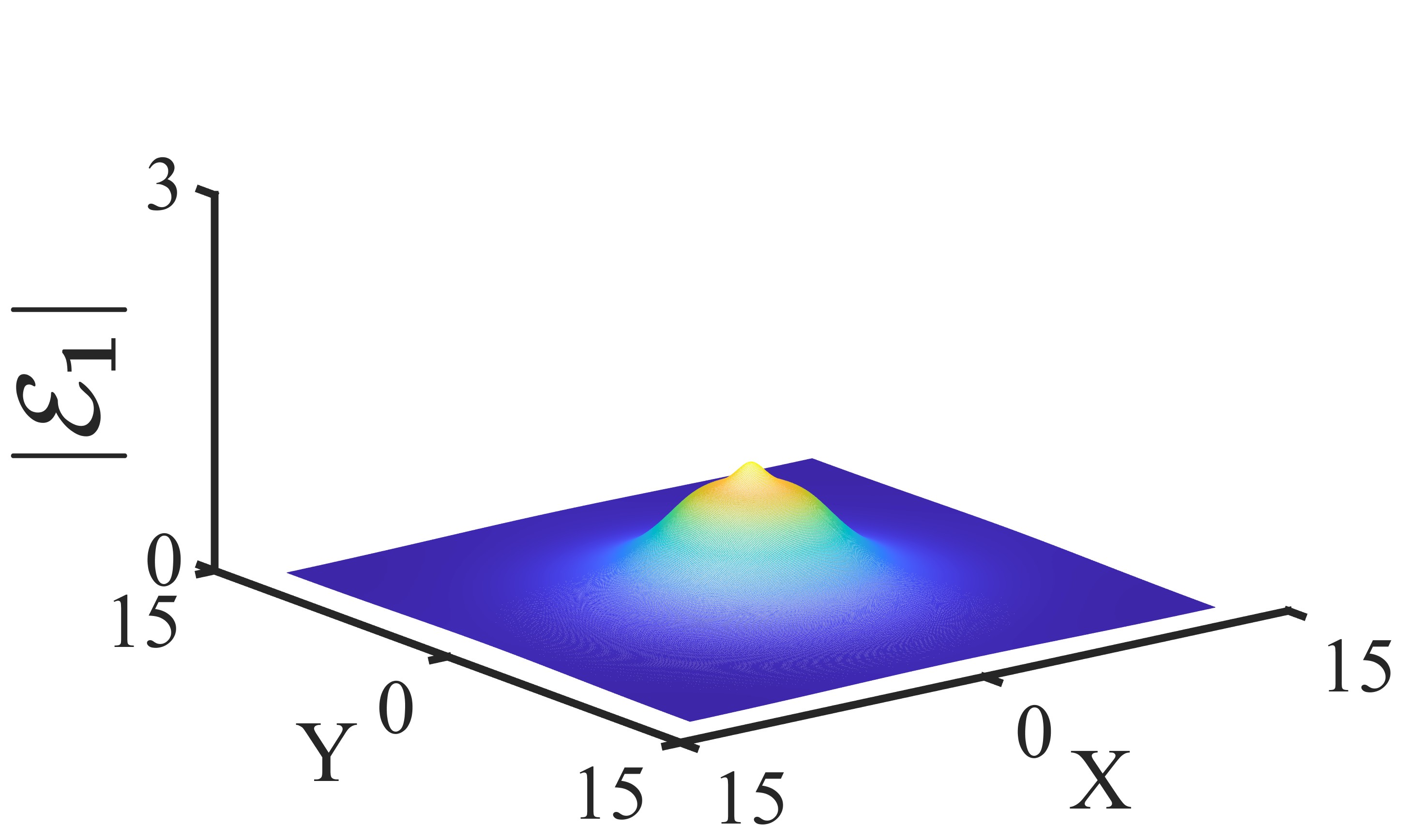}}}} \quad
      \subfigure[]{\scalebox{0.5}
{\includegraphics[width=3in,height=2in]{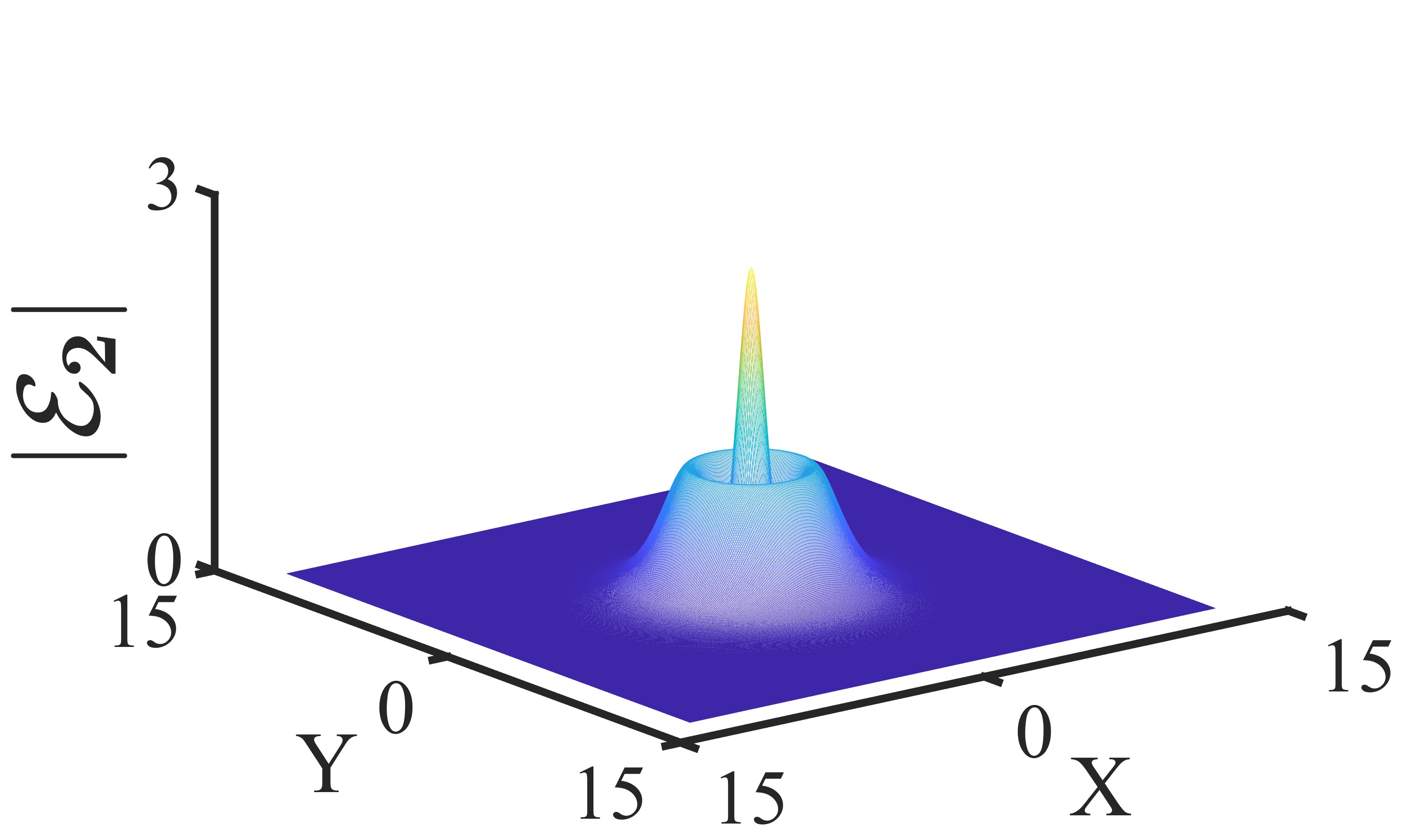}}}
      \subfigure[]{\scalebox{0.5}
{\includegraphics[width=3in,height=2in]{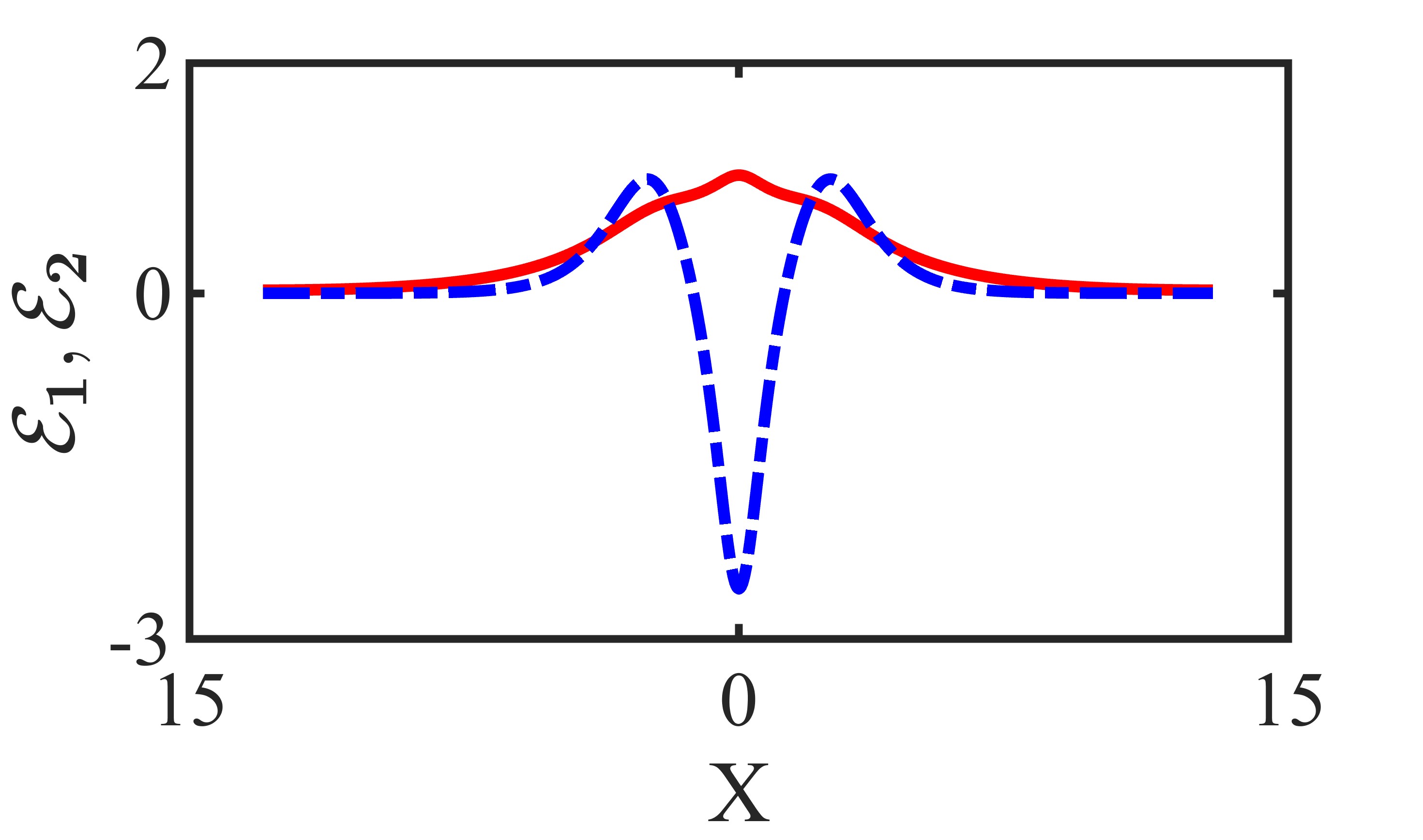}}}
	  \subfigure[]{\scalebox{0.5}
{\includegraphics[width=3in,height=2in]{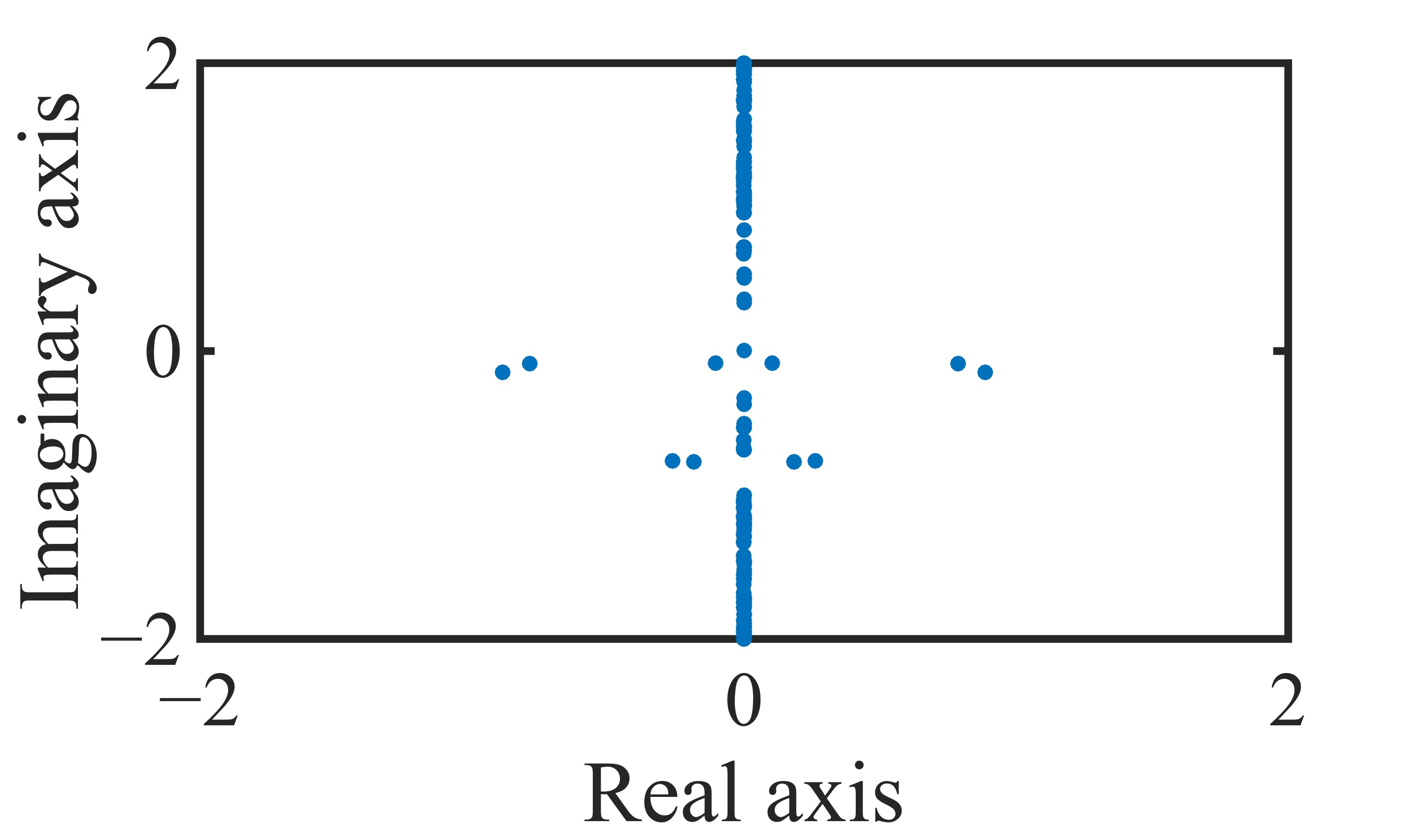}}} \quad
    \caption{3D profile of field $|{\cal E}_1(x,y)|$ with power  $P_1=5.0224$. (b) 3D profile of field $|{\cal E}_2(x,y)|$with power $P_2=5.4583$. (c) Radial profile of both electric fields ${\cal E}_1(x,y)$ (red) and ${\cal E}_2(x,y)$ (blue) with total power $P=P_1+P_2=10.4807$. (d) Linear stability spectrum}
    \label{soliton5}
  \end{center}
\end{figure}

\begin{figure}[htbp]
  \begin{center}
    \mbox{
      \subfigure[]{\scalebox{0.5}{\includegraphics[width=3in,height=2in]{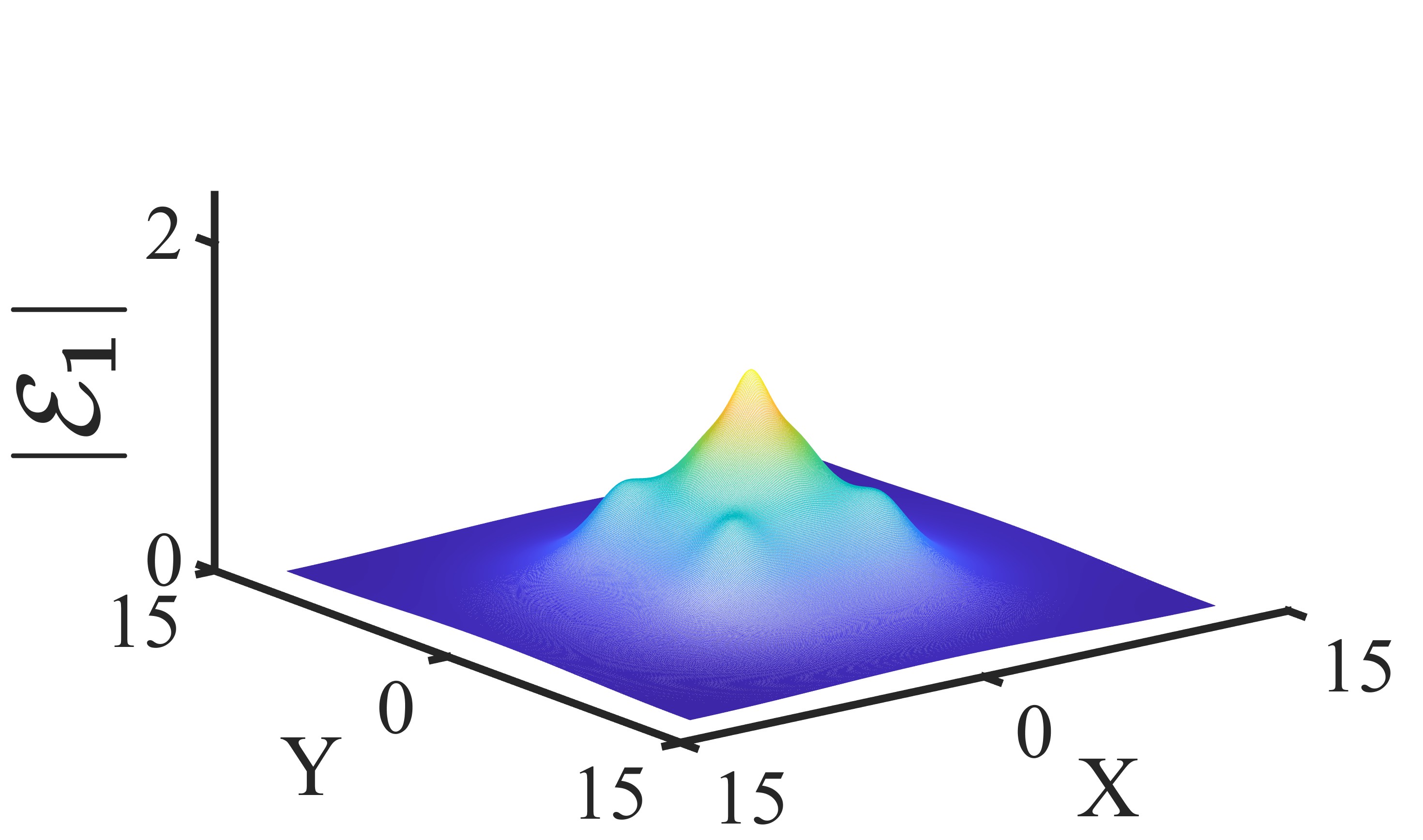}}}} \quad
      \subfigure[]{\scalebox{0.5}
{\includegraphics[width=3in,height=2in]{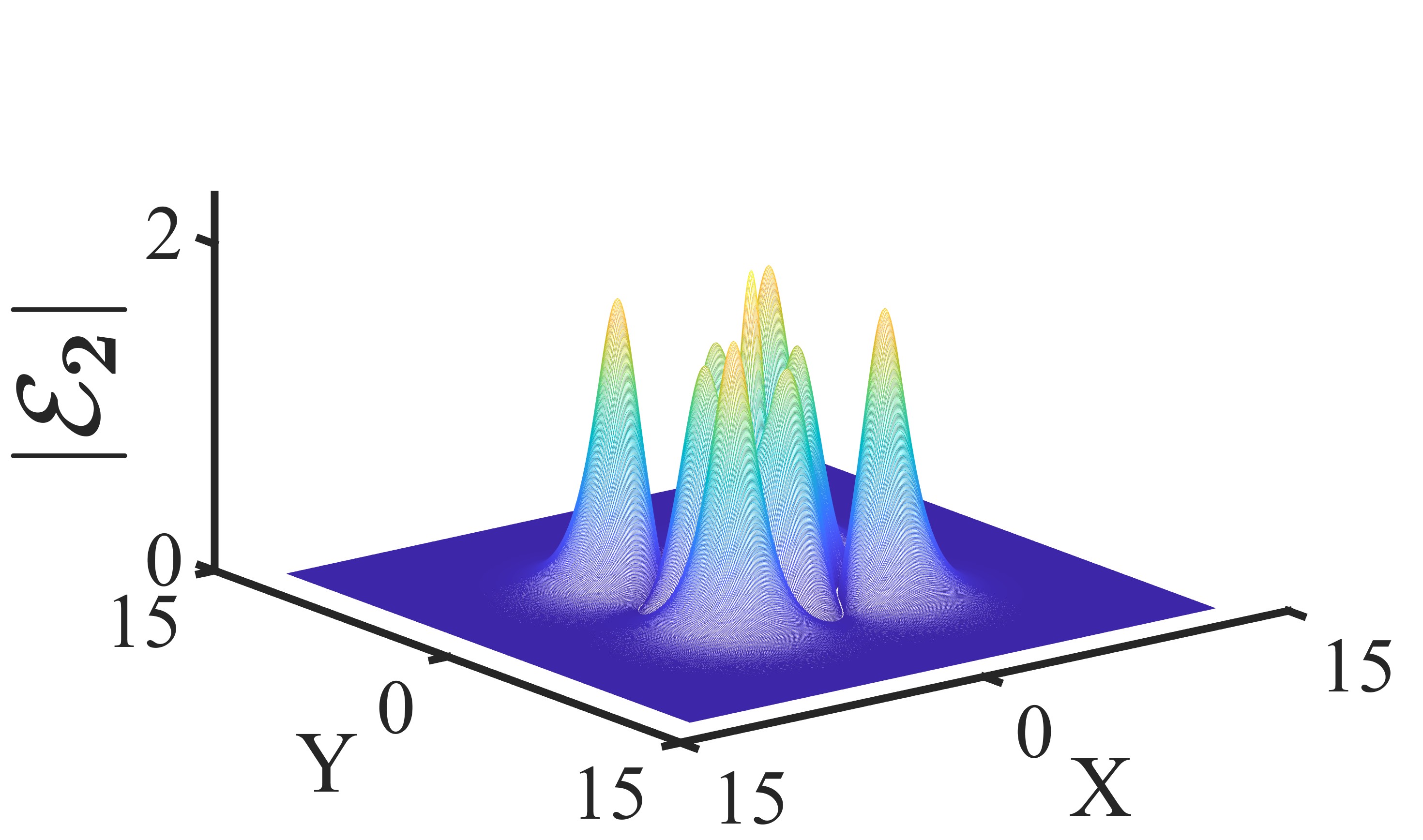}}}
      \subfigure[]{\scalebox{0.5}
{\includegraphics[width=3in,height=2in]{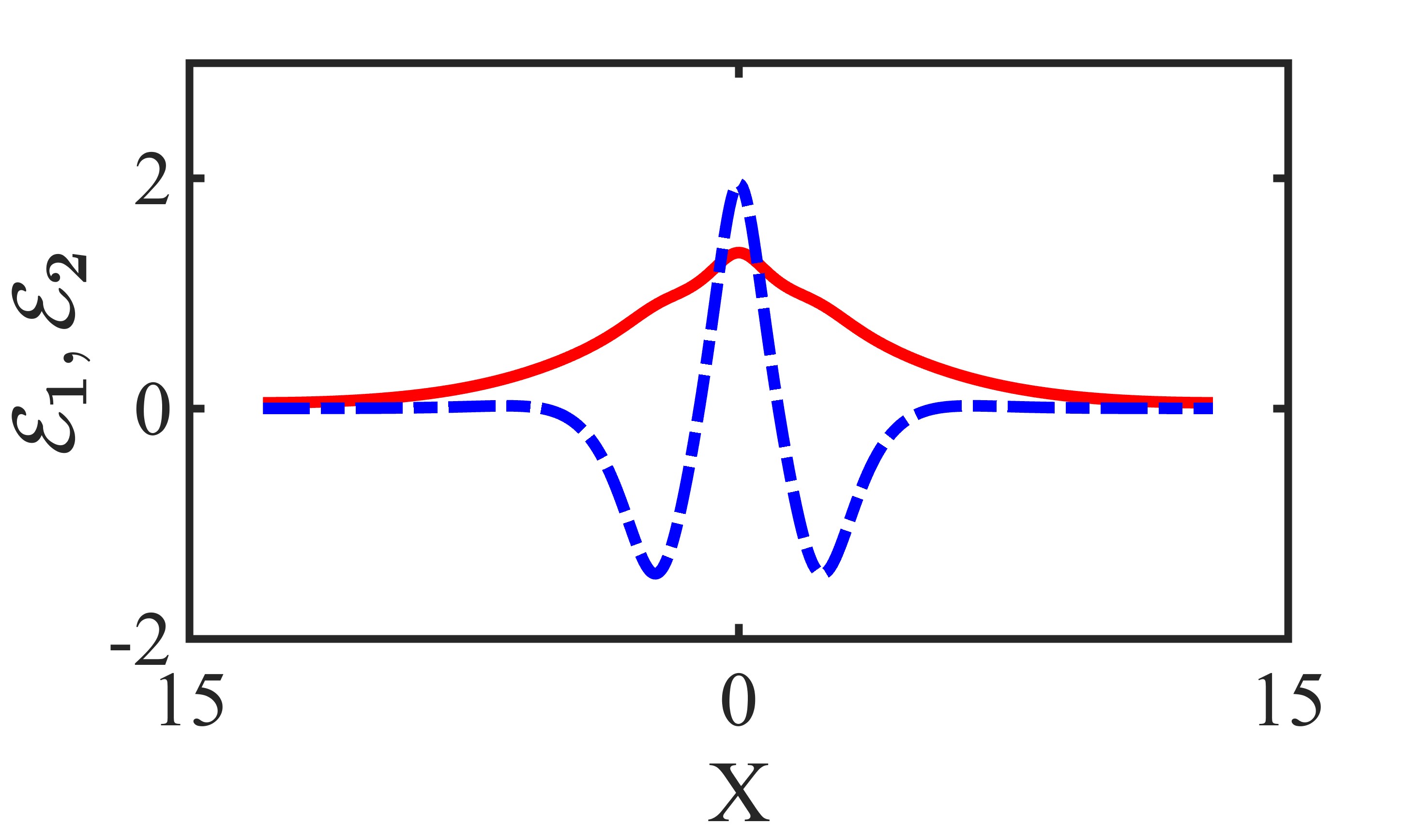}}}
	  \subfigure[]{\scalebox{0.5}
{\includegraphics[width=3in,height=2in]{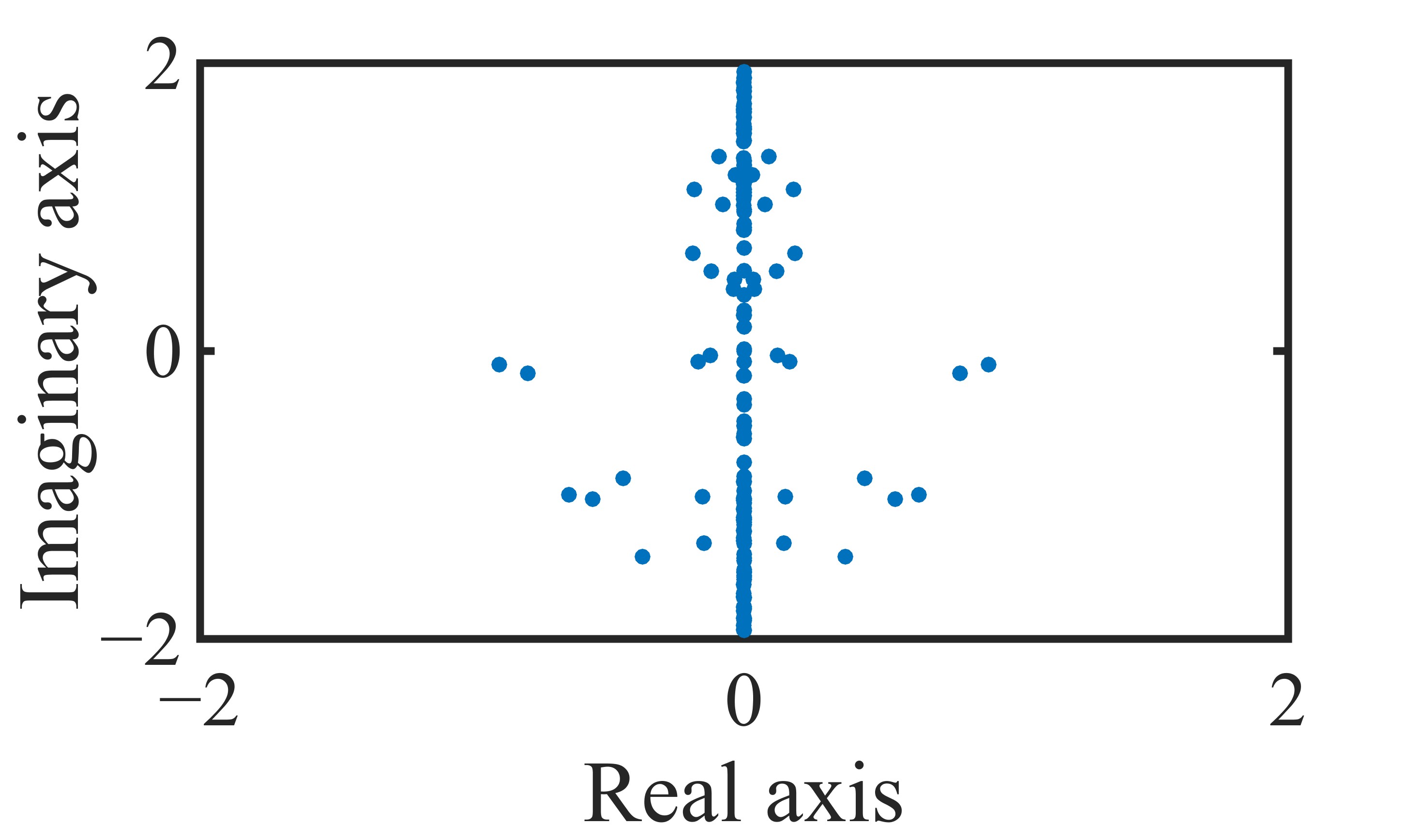}}} \quad
    \caption{3D profile of field $|{\cal E}_1(x,y)|$ with power  $P_1=9.3549$. (b) 3D profile of field $|{\cal E}_2(x,y)|$with power $P_2=10.0388$. (c) Radial profile of both electric fields ${\cal E}_1(x,y)$ (red) and ${\cal E}_2(x,y)$ (blue) with total power $P=P_1+P_2=19.3937$. (d) Linear stability spectrum}    
    \label{soliton6}
  \end{center}
\end{figure}

The system (\ref{bvp1},\ref{bvp2}) is solved numerically using the Newton-Conjugate-Gradient method \cite{yang2009newton}. Figures~\ref{soliton1}--\ref{soliton6} show representative two-color localized states $({\cal E}_1,{\cal E}_2)$ obtained for different powers. For each solution, we highlight both the full two-dimensional intensity profiles and the corresponding radial cross-sections, computed on a $512\times512$ grid. To characterize these states, we consider the individual component powers. For example, the solution shown in Fig.~\ref{soliton1} has powers $P_1=\frac{1}{2\pi}\int|{\cal E}_1|^2dxdy=15.9753$ and $P_2=\frac{1}{2\pi}\int|{\cal E}_2|^2dxdy=2.3873$. Throughout this work, the total power is defined as
$P=\frac{1}{2\pi}\int\left(|{\cal E}_1|^2+|{\cal E}_2|^2\right)dxdy$. 

Based on the relative power distribution between the two harmonics, we classify these solutions into two families. The first family (Figs.~\ref{soliton1}--\ref{soliton4}) is "fundamental-dominated", with the fundamental component carrying the larger share of the total power ($P_1>P_2$). The second family (Figs.~\ref{soliton5}-\ref{soliton6}) is "third-harmonic-dominated", for which the third harmonic carries the larger share of the total power ($P_2>P_1$). Linear stability properties of these states are shown in Figs.~\ref{soliton1}(d)--\ref{soliton6}(d), which display the spectra of the corresponding linearized operators. In every case, the spectrum contains eigenvalues with positive real parts, demonstrating that all stationary solutions are linearly unstable. 

\vspace{-4mm}
\vspace{-4mm}
\vspace{-4mm}

\subsection*{Collapse Events}

To investigate the dynamical behavior of these resonantly coupled unstable nonlinear modes, we perform direct numerical simulations of Eqs. (\ref{main1},\ref{main2}) using perturbed stationary states as initial conditions. All simulations are carried out using a pseudo-spectral Fourier method with spatial resolutions up to $512 \times 512$ points.

\begin{figure}[htbp]
  \begin{center}
    \mbox{
      \subfigure[]{\scalebox{0.5} {\includegraphics[width=3in,height=2in]{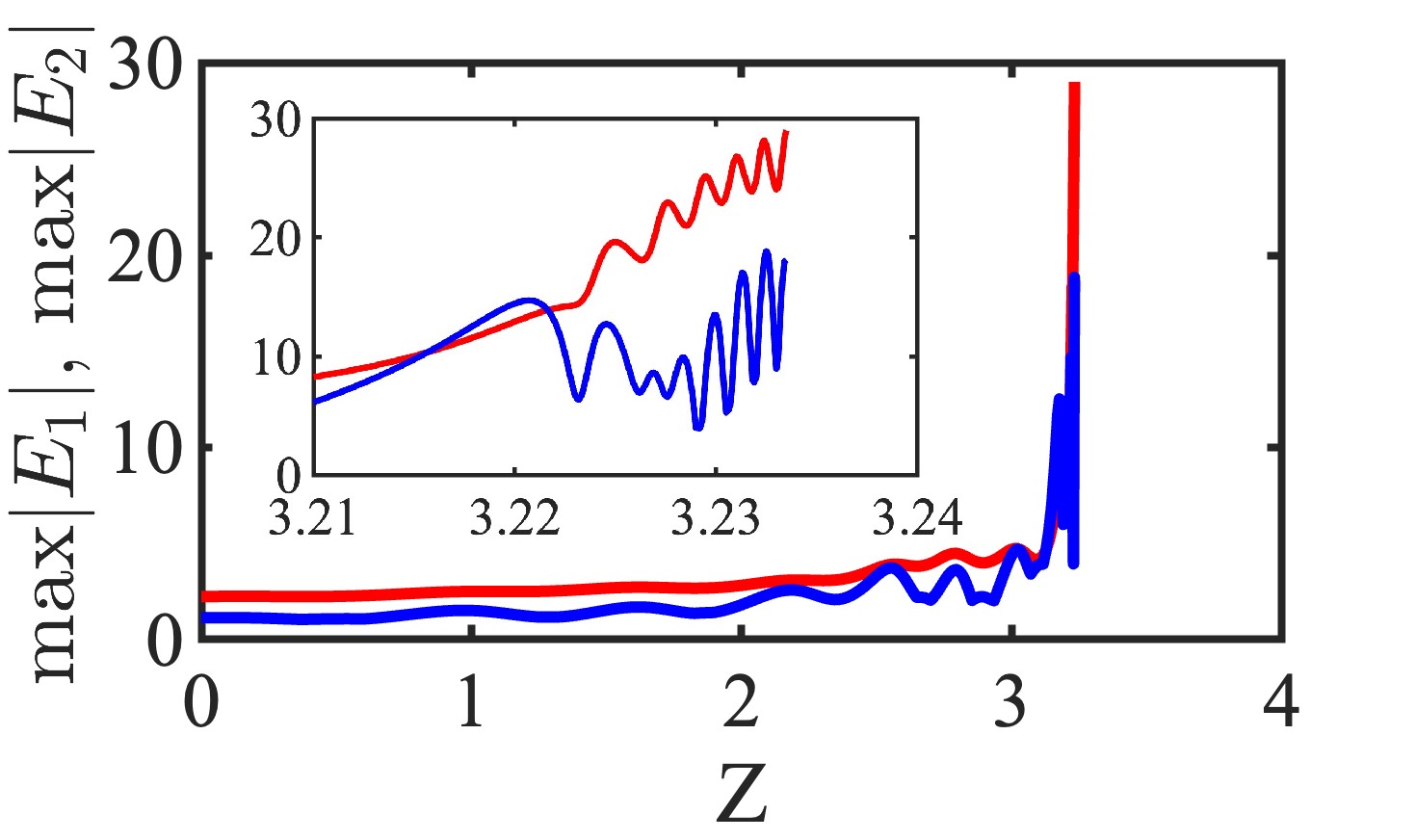}}}} \quad
      \subfigure[]{\scalebox{0.5}
{\includegraphics[width=3in,height=2in]{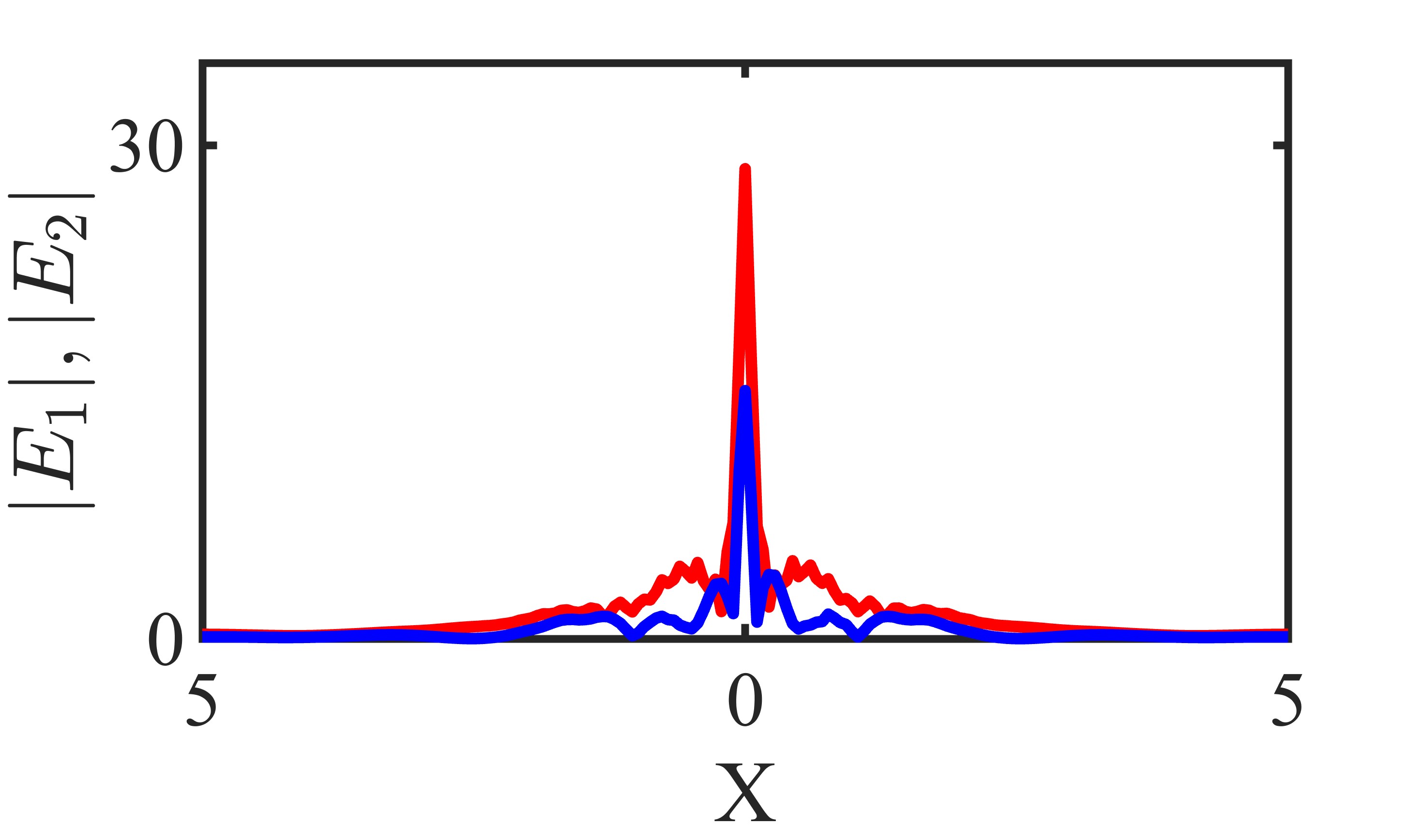}}}\quad
      \subfigure[]{\scalebox{0.5}
{\includegraphics[width=3in,height=2in]{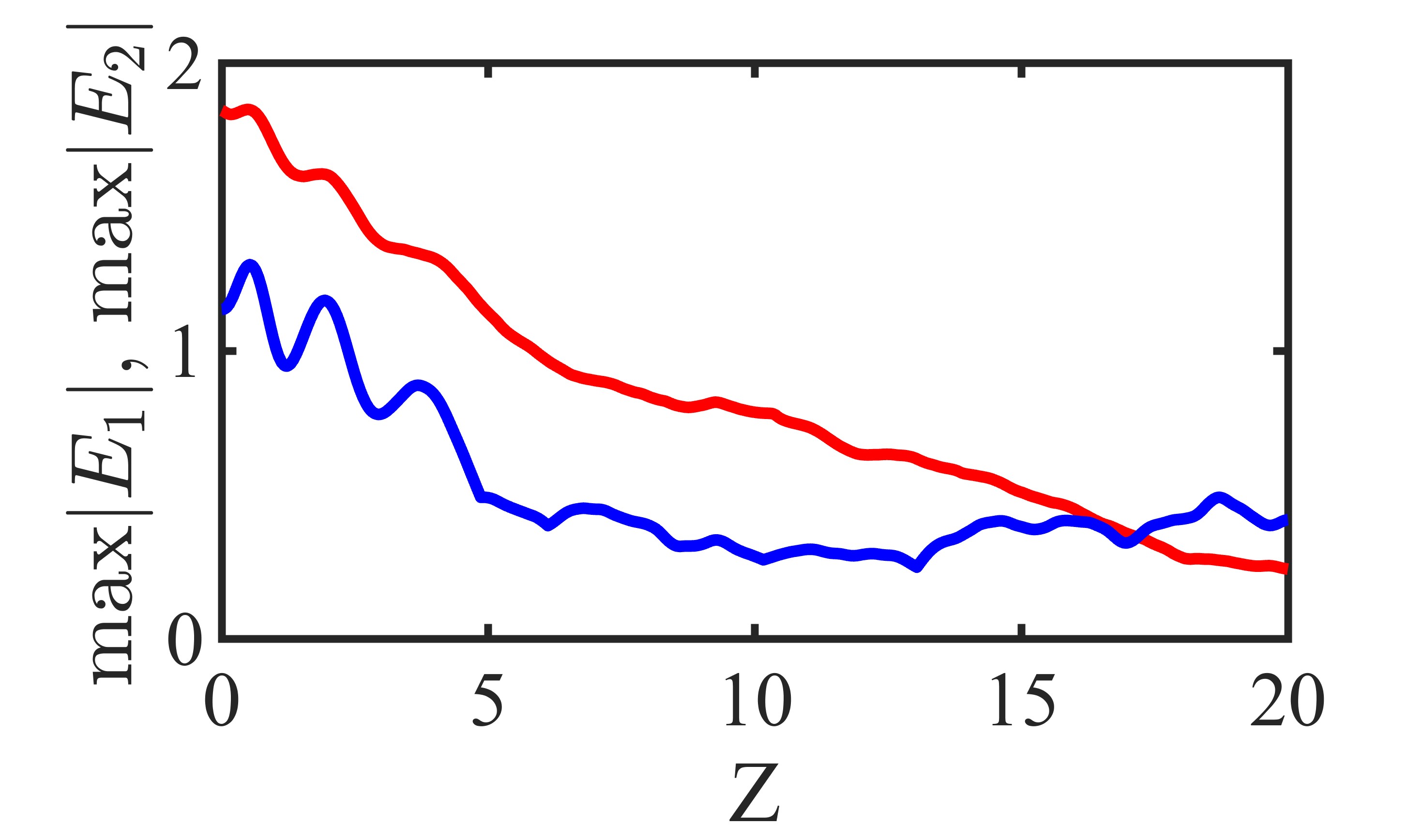}}}\quad
      \subfigure[]{\scalebox{0.5}
{\includegraphics[width=3in,height=2in]{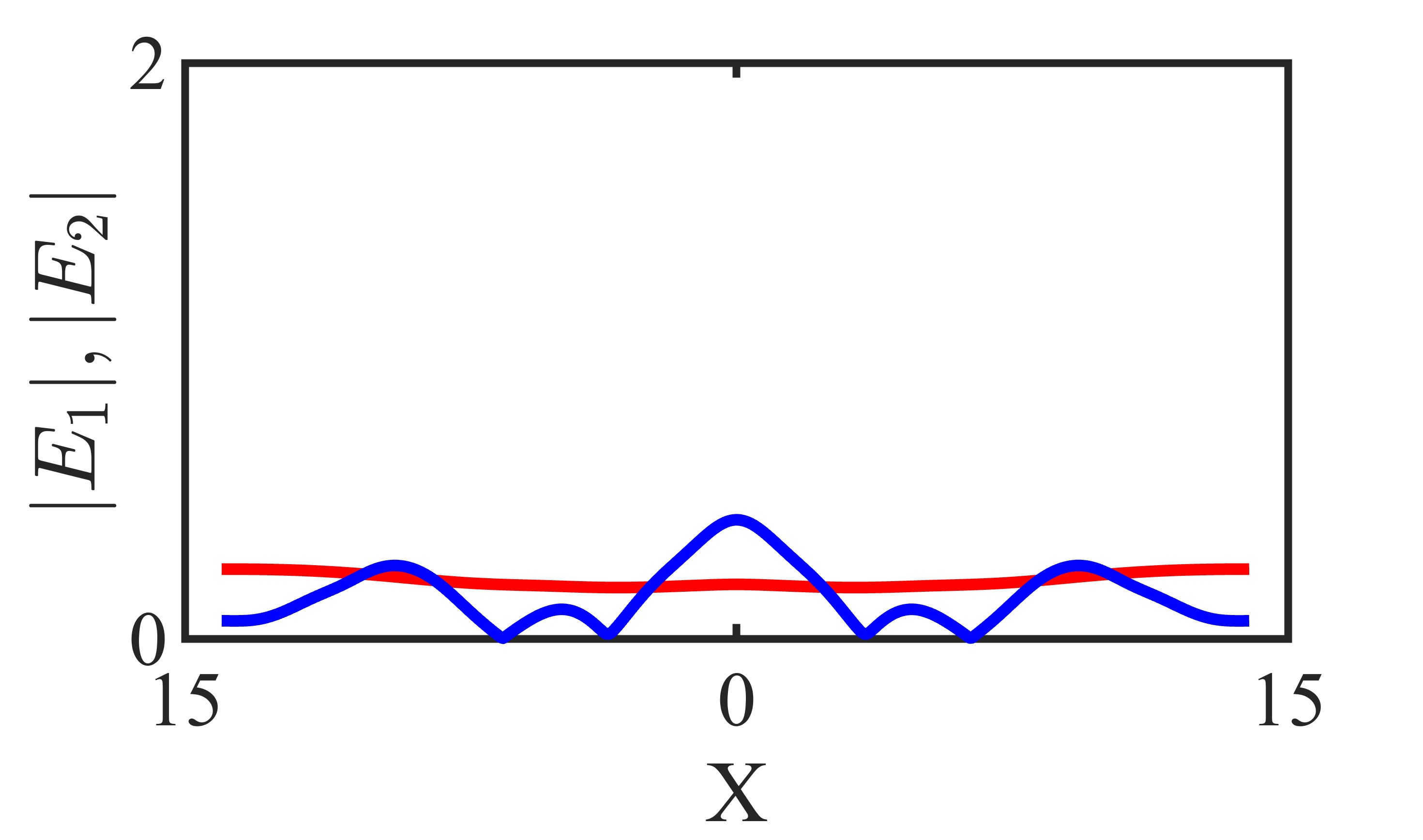}}}\quad
    \caption{(a) Propagation of peak amplitudes of perturbed two color soliton shown on Figure \ref{soliton1} with individual powers  $P^{+}_1=19.3301$ (red),$\,\,\,P_2=2.3873$ (blue), (b) Radial profiles of fields at distances $z=3.2335$, (c) Propagation of peak amplitudes of perturbed two color soliton shown on Figure \ref{soliton1} with individual powers $\,\,\,P^{-}_1=12.94$ (red),$\,\,\,P_2=2.3873$ (blue), (d) Radial profiles of fields at distances $z=20$}
    \label{soliton1_collapse}
  \end{center}
\end{figure}

We first consider the lowest-power stationary state of the fundamental-dominated family, shown in Fig.~\ref{soliton1}, and perturb its amplitude. Increasing the total power by perturbing only the fundamental component, $E_1=1.1{\cal E}_1$ and $E_2={\cal E}_2$, leads to simultaneous self-focusing of both fields. As shown in Fig.~\ref{soliton1_collapse}(a), the peak amplitudes of both components grow rapidly, indicating collapse-like dynamics. Conversely, reducing the total power by setting $E_1=0.9{\cal E}_1$ while keeping $E_2={\cal E}_2$ causes both beams to diffract, with their peak amplitudes decaying monotonically during propagation, as illustrated in Fig.~\ref{soliton1_collapse}(c).

\begin{figure}[htbp]
  \begin{center}
    \mbox{
      \subfigure[]{\scalebox{0.5} {\includegraphics[width=3in,height=2in]{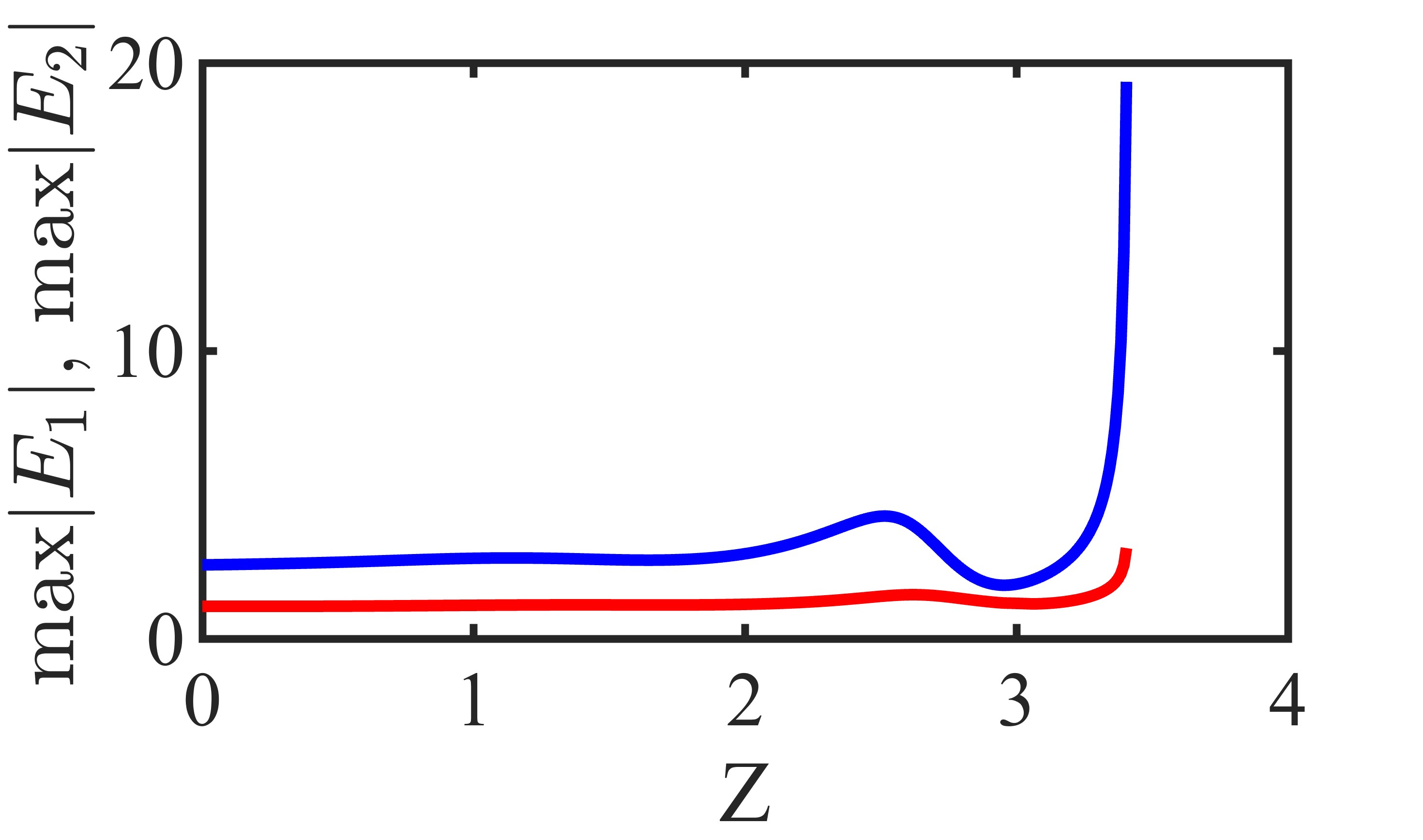}}}} \quad
      \subfigure[]{\scalebox{0.5}
{\includegraphics[width=3in,height=2in]{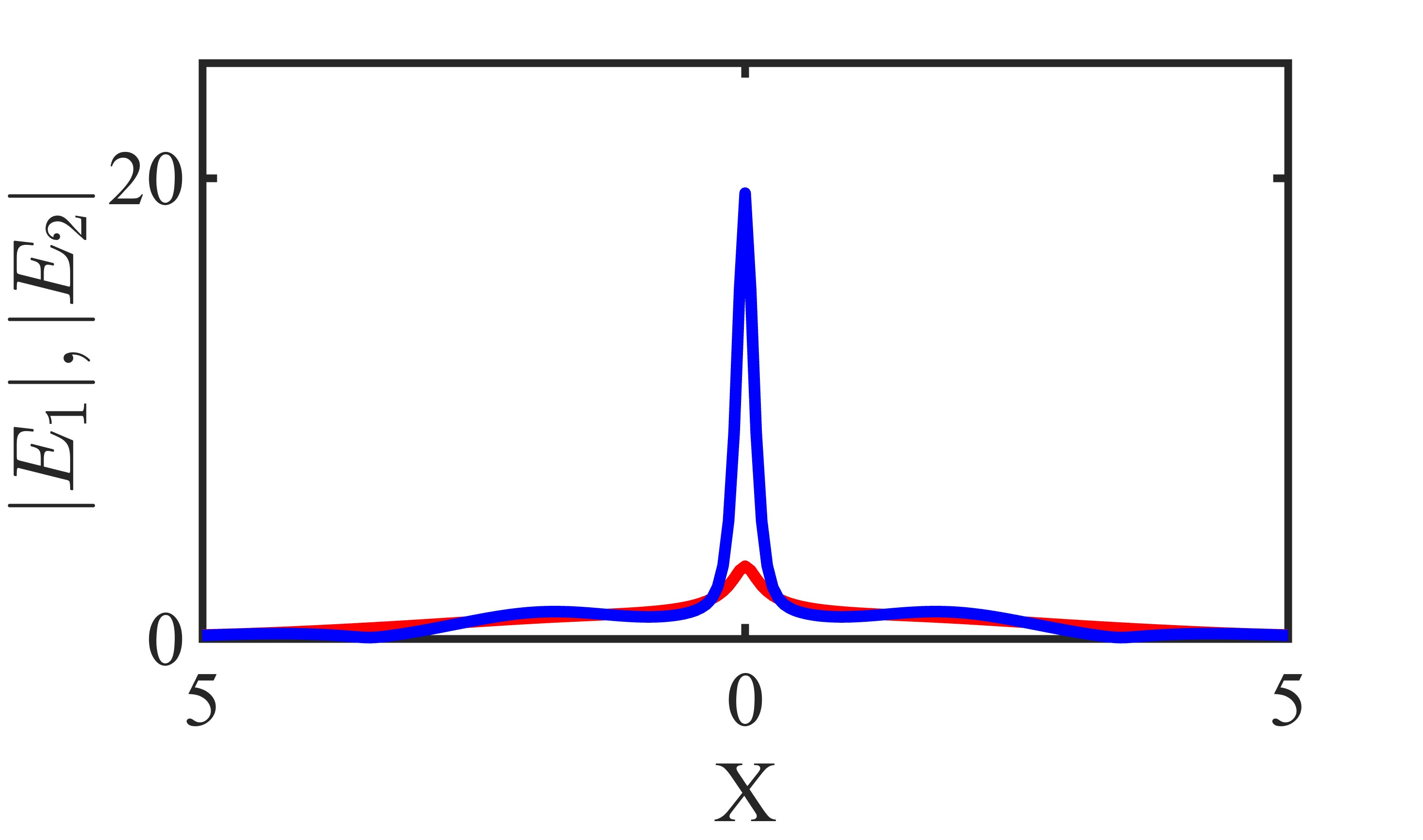}}}\quad
      \subfigure[]{\scalebox{0.5}
{\includegraphics[width=3in,height=2in]{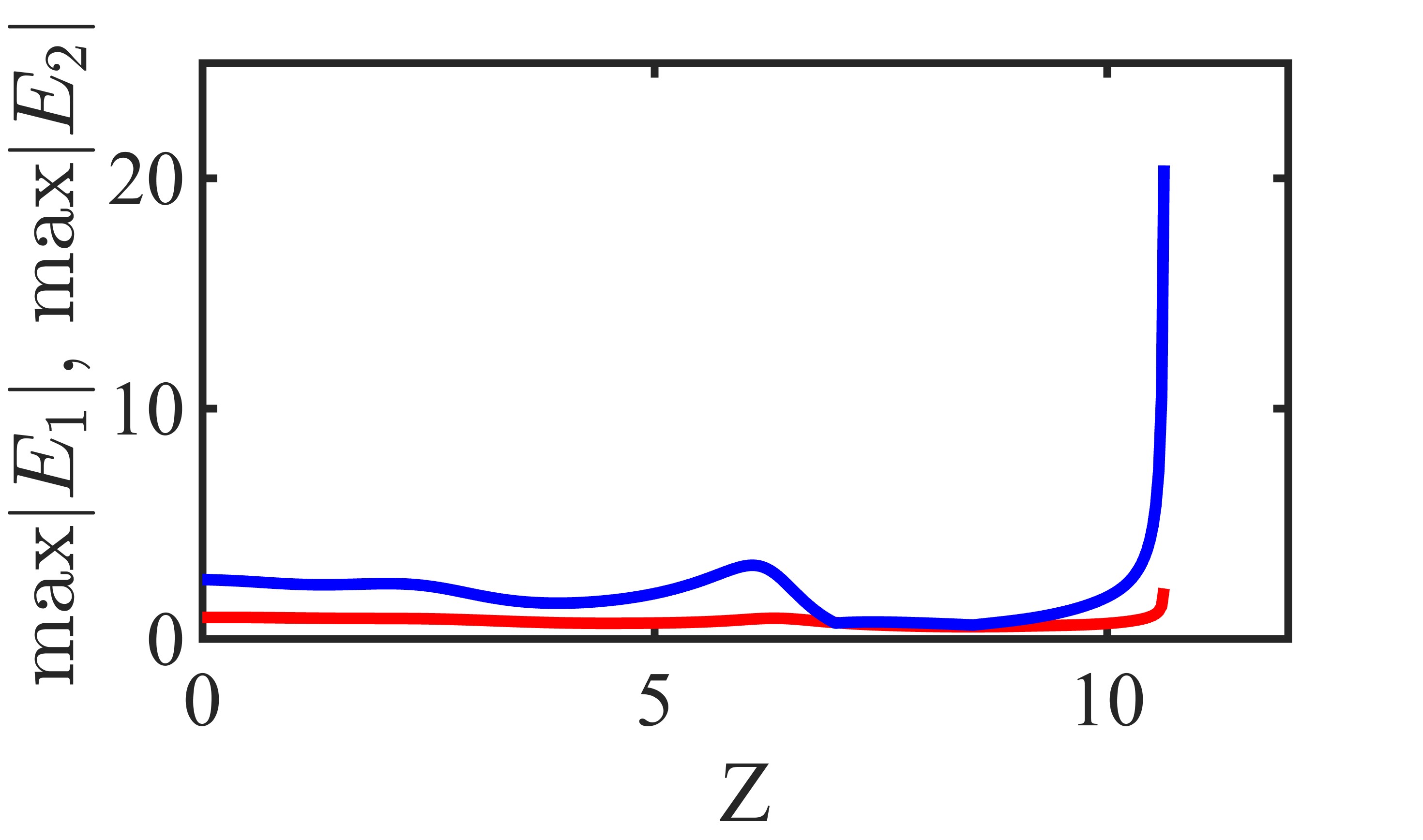}}}\quad
      \subfigure[]{\scalebox{0.5}
{\includegraphics[width=3in,height=2in]{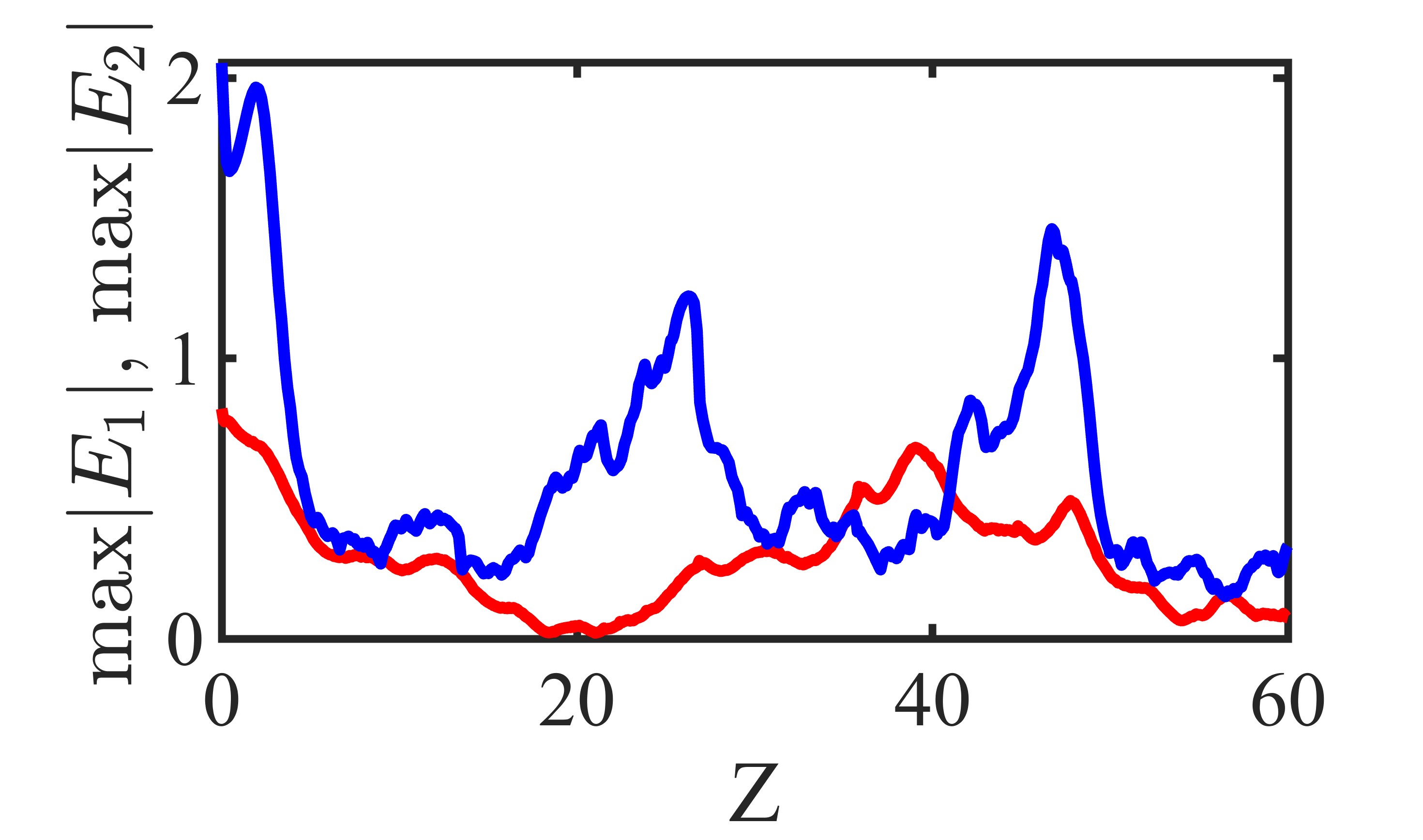}}}\quad
    \caption{(a) Propagation of peak amplitudes of perturbed two color soliton shown on Figure \ref{soliton5} with individual powers $P^{+}_1=6.0771$ (red),$\,\,\,P_2=5.4583$ (blue), (b) Radial profiles of fields at distances $z=3.405$, (c) Propagation of peak amplitudes of perturbed two color soliton shown on Figure \ref{soliton5} with individual powers $P^{-}_1=4.0681$ (red),$\,\,\,P_2=5.4583$ (blue), (d) Propagation of peak amplitudes of perturbed two color soliton shown on Figure \ref{soliton5} with individual powers $P^{-}_1=3.2143$ (red),$\,\,\,P^{-}_2=3.4933$ (blue)}
    \label{soliton2_collapse}
  \end{center}
\end{figure}

We observe qualitatively similar dynamics when perturbing higher-power stationary state, such as the solution shown in Fig. 2. A key feature emerging from the collapsing simulations (Figs. 7(a)) is the appearance of pronounced oscillations in the third-harmonic amplitude. These oscillations grow in magnitude as the system evolves and precede strong self-focusing. We interpret this behavior as a precursor to {\bf resonant two-color collapse}, reflecting the underlying energy exchange enabled by the phase-matched coupling.

To examine the dynamical role of the "third-harmonic-dominated" family, we repeat the propagation study using the lowest-power stationary solution shown in Fig. 5 as the initial condition. Perturbations of this solution do not reveal a clear threshold separating collapse from diffraction. Increasing the amplitude of the fundamental component, ($E_1=1.1\mathcal{E}_1$), while keeping the third harmonic unchanged, leads to simultaneous self-focusing of both components, as shown in Fig. 8(a,b). In contrast to the dynamics observed in Fig. 7, the evolution remains essentially monotonic, with neither component exhibiting the pronounced oscillations that characterize resonant collapse in the "fundamental-dominated" family. Similar non-oscillatory behavior is also observed when the amplitude is reduced to ($E_1=0.9\mathcal{E}_1$), indicating that the collapse dynamics are qualitatively different for this family of stationary states. When the amplitudes of both components are reduced further, ($E_1=0.8\mathcal{E}_1$) and ($E_2=0.8\mathcal{E}_2$), the nonlinear interaction is no longer sufficient to balance diffraction, and both beams broaden during propagation, as illustrated in Fig. 8(d). These results suggest that the "third-harmonic-dominated" stationary states do not act as "dynamical separatrices" between collapse and diffraction in the same manner as the "fundamental-dominated" family. Instead, moderate perturbations on either side of the stationary solution may evolve toward collapse, while sufficiently large reductions of the total power eventually lead to diffraction. The absence of resonant oscillations together with the lack of a well-defined collapse–diffraction boundary indicates that the two stationary families play fundamentally different dynamical roles, reinforcing the conclusion that no universal critical power exists in the resonantly coupled third-harmonic generation model.

\subsection*{Conclusions}

In summary, we have investigated a third-harmonic generation model in a Kerr medium and analyzed the dynamics of resonantly coupled beams at frequencies  $\omega_2=3\omega_1$. Our results provide evidence for qualitatively new self-focusing behavior in comparison with the standard (2+1)D nonlinear Schrödinger equation. In particular, we identified linearly unstable two-color localized states that act as structures separating regimes of simultaneous self-focusing and joint diffraction.

Unlike the single-component case, where collapse is rigorously established via the virial theorem, the present system does not readily admit such analysis. While our numerical simulations strongly suggest collapse-like dynamics, we do not claim a rigorous proof of singularity formation. Instead, our results point to a more intricate scenario in which the evolution is governed by the distribution of power between the harmonics, and where the notion of a universal critical power may not apply.

We further observe that the two components can exhibit distinct collapse rates and that the third-harmonic field develops pronounced oscillatory behavior preceding strong self-focusing. These features indicate that resonant coupling fundamentally alters the approach to collapse.

Our findings raise several open questions, including the existence of self-similar solutions near singularity formation, the possibility of defining generalized collapse criteria, and the precise dynamical role of the two-color stationary states. Addressing these issues will be essential for developing a complete theory of multi-frequency self-focusing in Kerr media.
While advancing understanding on the theoretical front may require adding more physics to the model, we believe these first results already provide useful guidelines for experimental efforts, to what are best initial scenarios to launch in the formation of co-existing 
UV/IR light filaments. 


\bibliography{ResonantSelfFocusing}

\end{document}